\documentclass[a4paper,twocolumn]{IEEEtran}
\usepackage[latin9]{inputenc}
\usepackage{units}
\usepackage{amsmath}
\usepackage{amssymb}
\usepackage{graphicx}

\makeatletter

\pdfpageheight\paperheight
\pdfpagewidth\paperwidth

\usepackage{cite}

\makeatother

\begin{document}
\title{New Lower Bounds on the Capacity of Optical Fiber Channels via Optimized
Shaping and Detection}
\author{Marco Secondini \IEEEmembership{Senior Member, IEEE}, Stella Civelli
\IEEEmembership{Member, IEEE}, Enrico Forestieri \IEEEmembership{Senior Member, IEEE},
Lareb Zar Khan\thanks{M. Secondini, S. Civelli, E. Forestieri, and L. Zar Khan are with
the Institute of Communication, Information and Perception Technologies,
Scuola Superiore Sant'Anna, Pisa, Italy, and with the National Laboratory
of Photonic Networks, CNIT, Pisa, Italy (email: marco.secondini@sssup.it;
stella.civelli@sssup.it; forestieri@sssup.it; larebzar.khan@santannapisa.it).
This paper was presented in part at the European Conference on Optical
Communication, Bordeaux, France, September 2021 \cite{Secondini:2021ECOC}.}}
\maketitle
\begin{abstract}
Constellation shaping is a practical and effective technique to improve
the performance and the rate adaptivity of optical communication systems.
In principle, it could also be used to mitigate the impact of nonlinear
effects, possibly increasing the information rate beyond the current
limit dictated by fiber nonlinearity. However, this appealing idea
is frustrated by the difficulty of designing an effective shaping
strategy that takes into account the nonlinearity and long memory
of the fiber channel, as well as the possible interplay with other
nonlinearity mitigation strategies. As a result, only little progress
has been made so far, while the optimal shaping distribution and the
ultimate channel capacity remain unknown. In this work, we describe
a novel technique to optimize the shaping distribution in a very general
setting and high-dimensional space. For a simplified block-memoryless
nonlinear optical channel, the capacity lower bound obtained by the
proposed technique can be expressed analytically, establishing the
conditions for an unbounded growth of capacity with power. In a more
realistic scenario, the technique can be implemented by a rejection
sampling algorithm driven by a suitable cost function, and the corresponding
achievable information rate estimated numerically. The combination
of the proposed technique with an improved (non-Gaussian) decoding
metric yields a new capacity lower bound for the dual-polarization
WDM channel.
\end{abstract}

\begin{IEEEkeywords}
Optical fiber communication, channel capacity, achievable information
rate, nonlinearity mitigation, constellation shaping. 
\end{IEEEkeywords}

\section{Introduction}

Modeling and mitigation of fiber nonlinearity have been an important
aspect of optical fiber communication since the very beginning \cite{nl-agrawal}.
In the last decade, the revolution of coherent detection aided by
digital signal processing (DSP) has further stimulated the research
in this field, bringing two new elements to the picture: an almost
unlimited capability to modulate, demodulate, and process optical
signals by DSP; and the apparent impossibility to push the spectral
efficiency of optical systems beyond an alleged limit set by fiber
nonlinearity \cite{Ellis10,Roadmap2016,secondini_JLT2017_scope}.
Nonetheless, many problems related to fiber nonlinearity\textemdash optimal
modulation, optimal detection, and channel capacity to name a few\textemdash are
still open and actively investigated.

The problem of mitigating nonlinear effects and improving system performance
has been addressed from many different perspectives. One such perspective
is that of considering the optical fiber channel as a fixed element
of the system, whose performance can be improved only by deploying
proper DSP techniques at the transmitter or receiver. In this context,
a fundamental problem is the study of the capacity of the optical
fiber channel and the implementation of systems that operate at rates
close to it. This problem has been widely addressed in the literature,
considering different bounding techniques, fiber links, and configurations
\cite{splett1993ultimate,mitra:nature,Tang:JLT02,Turitsyn03,Kahn04,Djordjevic05,Taghavi06,Essiambre:JLT0210,Ellis10,yousefi11,polito:oexp,Mecozzi:JLT0612,dar:OL2014,Agrell14,agrell:tcom2015,Kramer2015,Agrell:jlt2015,agrell17ecoc,keykhosravi17,secondini_JLT2017_scope,secondini:ecoc17,secondini2019JLT,garcia2020mismatched1,garciagomez2021mismatched,sefidgaran:2021arxiv}.
For a more detailed analysis of the existing literature, we refer
to \cite{secondini_JLT2017_scope,SECONDINI2020867}. In this work,
we focus on single-mode fibers and wavelength-division multiplexing
(WDM) systems, studying the capacity from a single-user perspective
\cite{Agrell:jlt2015}. On the one hand, the tightest available capacity
lower bounds for this channel can be found by the methodologies proposed
in \cite{secondini2019JLT,garciagomez2021mismatched}, which yield
an achievable information rate (AIR) that reaches a peak at some optimum
power and then decays. On the other hand, the only available upper
bound equals the capacity of the additive white Gaussian noise (AWGN)
channel and hence increases indefinitely with power \cite{Kramer2015}.
The combination of these bounds does not rule out neither the existence
of a finite capacity limit nor the possibility of an unbounded growth
with power.

The capacity problem can be formulated in a way that entails two fundamental
subproblems: the optimization of the input distribution, which is
related to the implementation of an optimal coded modulation scheme;
and the optimization of the decoding metric, which is related to the
implementation of an optimal receiver. The maximization of the AIR
obtained for a given input distribution and decoding metric yields
the channel capacity. For the AWGN channel, the solution is well known.
The optimal decoding metric, matched to the channel conditional distribution,
factorizes into the product of marginal Gaussian distributions; analogously,
the optimal input distribution factorizes into the product of identical
marginal distributions\textemdash Gaussian in the general case\cite{shannon48},
Maxwell\textendash Boltzmann (MB) if the symbols are constrained on
a given discrete alphabet \cite{kschischang1993optimal}; the resulting
AIR equals the channel capacity $C=\log_{2}(1+\mathrm{SNR})$ (for
an unconstrained input alphabet), where $\mathrm{SNR}$ is the signal
to noise ratio.

The picture is quite different for the nonlinear fiber channel. In
this case, the conditional distribution is unknown, so that the receiver
is optimized for an \emph{auxiliary channel}\textemdash an approximated
version of the true channel, for which the optimal decoding metric
is available. This approach is known as \emph{mismatched decoding}.
Often, for simplicity and in the absence of a suitable alternative,
an AWGN auxiliary channel is considered, so that the decoding metric
is still taken as the product of marginal Gaussian distributions.
The search for more accurate and mathematically tractable mismatched
channel models is the subject of current research \cite{Agrell:ITW2015}.
For instance, several models show that interchannel nonlinear interference
(NLI) includes relevant phase and polarization noise (PPN) components
that evolve slowly in time \cite{Sec:PTL2012,Dar2013:opex,dar_JLT2017_nonlinear}.
Such components depend also on frequency and can be alternatively
represented as time-varying linear intersymbol interference \cite{secondini2013achievable}.
Their mitigation is possible \cite{Sec:PTL14,dar_JLT2017_nonlinear}
and yields an increase of the AIR, which is more effective if combined
with subcarrier multiplexing \cite{Mar:OFC15,secondini2019JLT} and
an optimized per-subcarrier power allocation \cite{garcia2020mismatched1,garciagomez2021mismatched}.
Moreover, even the additive component of NLI has some correlation
in time, which might be exploited for its mitigation \cite{garcia2020mismatched1,garciagomez2021mismatched}.

Constellation shaping improves the efficiency of a digital modulation
scheme by modifying the position of the symbols in the constellation
diagram (geometric shaping) or the frequency with which they are used
(probabilistic shaping), trying to match the optimal input distribution
for the given channel. A practical coded modulation scheme that has
attracted much interest in recent years is probabilistic amplitude
shaping (PAS), thanks to its nearly optimal performance, simple implementation,
and fine rate granularity \cite{bocherer2015bandwidth,buchali2016JLT}.
PAS uses a distribution matcher, followed by a systematic forward
error correction (FEC) encoder, to induce the desired distribution
over a quadrature amplitude modulation (QAM) constellation. The optimal
condition of i.i.d. MB symbols is approached as the block length of
the distribution matcher goes to infinity \cite{bocherer2015bandwidth,gultekin2019TWireless}.

Constellation shaping can be used also to mitigate nonlinear effects.
In this case, often referred to as nonlinear constellation shaping,
the location or probability of the constellation symbols are optimized
to minimize the amount or impact of the generated NLI. The main problem
here is that the optimal input distribution is unknown, so that the
distribution matcher approach of PAS cannot be directly implemented.
There are many evidences suggesting that optimizing the marginal distribution
of i.i.d. 2D symbols yields negligible gains in this case \cite{fehenberger2016JLT}.
In fact, to unlock the full potentiality of nonlinear constellation
shaping, the optimization should be performed in a higher dimensional
space. So far, the approaches have been limited to the optimization
of low-rate constellations in a low-dimensional space (e.g., geometric
shaping in 4D and 8D \cite{kojima2017nonlinearity,chen2019polarization}),
or to a highly constrained optimization of PAS in a higher-dimensional
space (e.g., optimizing the block length of the distribution matcher
while keeping an MB target distribution) \cite{geller2016shaping,fehenberger2020mitigating,cho:2021preprint,gultekin:2021arxiv}.
The advantages obtained in this way are moderate, and might become
negligible in the presence of carrier recovery algorithms \cite{civelli2020interplayECOC}.

The current research challenge is the full optimization of the constellation
in a high-dimensional space, possibly in combination with improved
decoding strategies. While this is an extremely complex and still
unsolved problem, in this work we propose a capacity lower-bounding
technique based on rejection sampling to estimate the gain achievable
by such an optimization. The idea, named \emph{sequence selection},
was briefly introduced in \cite{civelli:2021ECOC}. In this work,
we better formalize the technique, we describe some practical procedures
for its implementation, and we use it to derive some new analytical
and numerical results on the optical fiber capacity.

The paper is organized as follows. Section~II introduces the optical
fiber channel and the capacity problem. Section~III describes the
proposed sequence selection technique and the related capacity lower
bound. Section~IV derives an analytical capacity lower bound for
a simplified block-memoryless optical channel, establishing the conditions
for an unbounded growth of capacity with power. Section~V presents
some numerical results and the new capacity lower bound for the WDM
channel. The conclusions are finally drawn in Section~VI.

\emph{Notation}: a random variable is denoted by an uppercase letter,
e.g., $X$, its expectation by $E\{X\}$, and its realization by the
corresponding lowercase letter $x$. This rule, however, is broken
when dealing with some deterministic quantities, e.g., the bandwidth
$W$, the power $P$, or the length $N$. The probability density
function (or \emph{distribution}, in short) of the random variable
$X$ is simply denoted by $p(x)$, with the argument implicitly defining
the specific distribution, so that $p(x)$ and $p(y)$ denote the
different distributions of the variables $X$ and $Y$. A subscript
is sometimes used to distinguish some specific distributions, e.g.,
the unbiased distribution $p_{u}(x)$. A discrete-time stochastic
process is denoted by a boldface uppercase letter, e.g., $\boldsymbol{X}=(X_{1},X_{2},\ldots)$,
and the notation $\boldsymbol{X}_{k}^{n}=(X_{k},X_{k+1},\ldots,X_{n})$
and $\boldsymbol{X}^{n}=(X_{1},X_{2},\ldots,X_{n})$ is used to denote
finite portions of the process, i.e., finite-length random sequences.
The corresponding realizations are denoted by boldface lowercase letters.
The energy of the sequence $\boldsymbol{x}^{n}$ is $\left\Vert \boldsymbol{x}^{n}\right\Vert ^{2}=\sum_{i=1}^{n}|x_{i}|^{2}$.
Integrals, if not otherwise specified by explicit limits, extend to
the whole space in which the integration variable is defined.

\section{System Description and Problem Formulation\label{sec:System-description-and}}

The goal of this work is to study the capacity of the optical fiber
channel, from the perspective of a single WDM user in a WDM configuration.
Here the optical fiber channel is defined as a waveform channel, in
which the propagation of the optical signal is governed by the Manakov
equation \cite{Wang99}. The Manakov equation accounts for attenuation,
dispersion, and Kerr nonlinearity, with the addition of a noise term
and a periodic gain/loss function that accounts for the presence of
optical amplifiers \cite{secondini2019JLT,SECONDINI2020867}.

The available optical bandwidth is divided into several independent
slots of size $W$, each allotted to a different WDM user. We assume
fair and independent WDM users, meaning that the multiplexed signals
are independently modulated by each user, have the same input power
and statistical properties, and are independently detected (behavioral
model \emph{c} in \cite{Agrell:jlt2015}). We assume that the input
and output signals of each user are strictly band-limited, with passband
bandwidth $W$. In this case, with no loss of generality, it is possible
to give an equivalent discrete-time formulation of the problem, where
the input and output signals are represented by their samples taken
at Nyquist rate $W$. The sampling theorem \cite{shannon48,proakis}
ensures that the input and output samples are a sufficient statistic
to represent the input and output waveforms, respectively, so that
any possible coding and decoding strategy for the waveform channel
can be implemented by working on the input and output samples. 
\begin{figure}
\begin{centering}
\includegraphics[width=1\columnwidth]{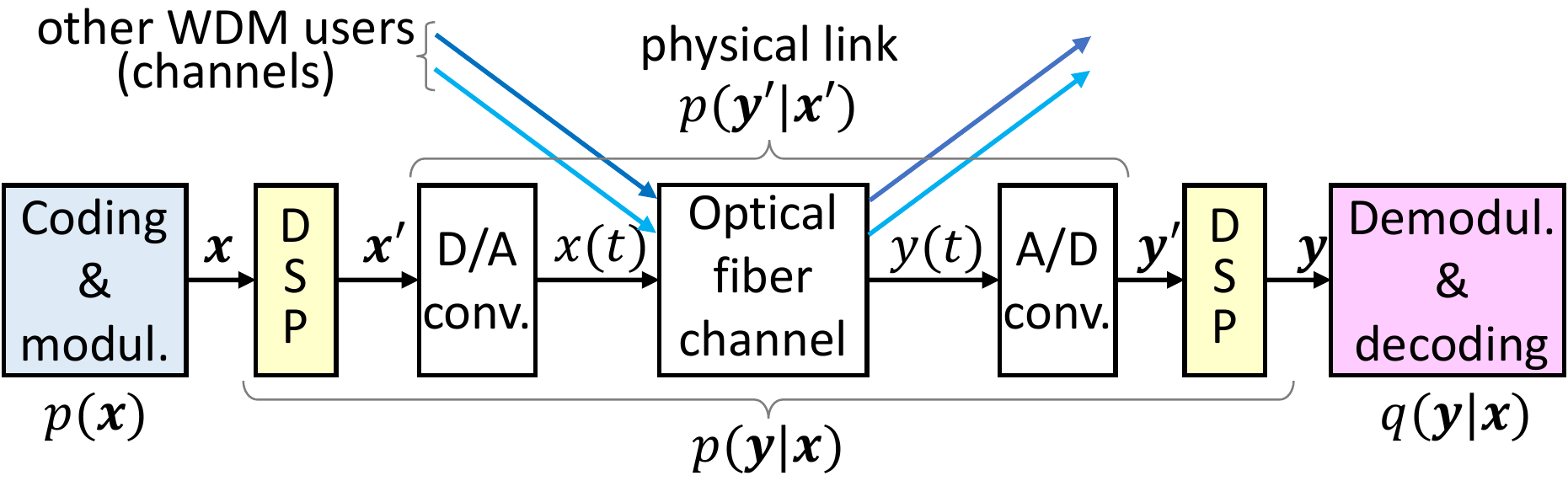}
\par\end{centering}
\caption{\label{fig:System-description}Description of the system considered
for AIR computation and maximization.}
\end{figure}

From a single-user perspective, the system is described as in Fig.~\ref{fig:System-description}.
At the transmitter, a random message (a sequence of independent uniformly
distributed bits) is encoded on a sequence of symbols $\boldsymbol{x}=(x_{1},x_{2},\ldots)$
by a proper combination of coding and modulation. The input sequence
$\boldsymbol{x}$ is processed by a digital signal processing (DSP)
block to obtain the sequence $\boldsymbol{x}'=(x_{1}',x_{2}',\ldots)$
and converted to the input waveform $x(t)$ by an ideal digital-to-analog
(D/A) converter with sampling rate and bandwidth $W$. After propagation
of $x(t)$ through the optical fiber channel, the output waveform
$y(t)$ is converted into the sequence $\boldsymbol{y}'$ by an ideal
analog-to-digital (A/D) converter with bandwidth and sampling rate
$W$ and digitally processed to obtain $\boldsymbol{y}$. The transmitted
message is finally recovered from $\boldsymbol{y}$ by demodulation
and decoding.

For every $N=1,2,\ldots$, the combination of coding and modulation
is characterized by the distribution $p(\boldsymbol{x}^{N})$ of the
emitted sequence; the discrete-time physical link\textemdash obtained
by the combination of D/A converter, optical fiber link, and A/D converter\textemdash is
characterized by the conditional distribution $p(\boldsymbol{y}'^{N}|\boldsymbol{x}'^{N})$;
and the combination of demodulation and decoding is characterized
by the decoding metric $q(\boldsymbol{y}^{N}|\boldsymbol{x}^{N})$.
Usually, DSP is employed at the transmitter and/or receiver to compensate
for some channel impairments (e.g., fiber dispersion) and to perform
some important tasks such as clock and carrier recovery. The combination
of DSP and discrete-time physical link gives the overall discrete-time
channel $p(\boldsymbol{y}^{N}|\boldsymbol{x}^{N})$.

The system performance is measured by the achievable information rate
(AIR) with mismatched decoding (also known as auxiliary-channel lower
bound) \cite{ArLoVoKaZe06}
\begin{equation}
I_{q}(\boldsymbol{X};\boldsymbol{Y})=\lim_{N\rightarrow\infty}\frac{1}{N}E\left\{ \log_{2}\frac{q(\boldsymbol{Y}^{N}|\boldsymbol{X}^{N})}{q(\boldsymbol{Y}^{N})}\right\} \label{eq:AIR-general}
\end{equation}
where
\begin{equation}
q(\boldsymbol{y}^{N})=\int q(\boldsymbol{y}^{N}|\boldsymbol{x}^{N})p(\boldsymbol{x}^{N})d\boldsymbol{x}^{N}\label{eq:mismatched_output_distribution}
\end{equation}
is the mismatched output distribution, obtained by connecting the
input source to an auxiliary channel with conditional distribution
$q(\boldsymbol{y}^{N}|\boldsymbol{x}^{N})$; the limit operation accounts
for channel memory; and the expectation is taken with respect to the
input distribution $p(\boldsymbol{x}^{N})$ and true channel distribution
$p(\boldsymbol{y}^{N}|\boldsymbol{x}^{N})$. In practice, (\ref{eq:AIR-general})
can be estimated by 
\begin{equation}
I_{q}(\boldsymbol{X};\boldsymbol{Y})\approx\frac{1}{N}\log_{2}\frac{q(\boldsymbol{y}^{N}|\boldsymbol{x}^{N})}{q(\boldsymbol{y}^{N})}\label{eq:AIR-general-MCestimate}
\end{equation}
taking $N$ sufficiently long to ensure that both the limit and expectation
operations are approximated with the desired accuracy.

The AIR (\ref{eq:AIR-general}) can be converted into a spectral efficiency
(SE) by dividing it by the product $WT$, where $W$ is the channel
bandwidth and $T$ the symbol rate. The choice $T=1/W$ made above
means that the AIR values obtained in this work (in bits/symbol) can
be directly read as SE values (in bits/s/Hz).

Single-mode fibers allow the propagation of two orthogonal polarization
modes. For the sake of simplicity, this is not explicitly considered
in Fig.~\ref{fig:System-description} and in the notation employed
in the paper, where symbols and signals are defined in $\mathbb{C}$
(2D space). In fact, when a WDM user exploits both modes\textemdash as
usual in modern coherent systems\textemdash the corresponding symbols
and signals should be considered as dual-polarization complex symbols
and signals, defined in the 4D space $\mathbb{C}^{2}$. In this case,
the AIR in (\ref{eq:AIR-general}) is expressed in bits/4D~symbol.
This specific scenario is considered in Sections~\ref{subsec:WDM-dual-polarization-system}
and \ref{subsec:Combination-of-shaping}, where the numerical results
are anyway reported in bits/2D~symbol and bits/s/Hz/polarization
by simply dividing the AIR by two.

The capacity problem studied in this work can be seen as an optimization
problem, where the AIR (\ref{eq:AIR-general}) of the system in Fig.~\ref{fig:System-description}
is maximized by optimizing the input distribution $p(\boldsymbol{x}^{N})$
(blue block), the decoding metric $q(\boldsymbol{y}^{N}|\boldsymbol{x}^{N})$
(pink block), and the DSP at transmitter and receiver (yellow blocks).
Any specific but suboptimal choice of these blocks yields a capacity
lower bound. In practice, this work and most of the current research
on this topic is focused on finding better combinations of these three
elements to improve on the existing lower bounds. The choice $T=1/W$
made above and the sampling theorem ensure that the capacity obtained
in this way equals the maximal SE for the waveform channel, no further
improvements being possible by changing the sampling rate.

With respect to the classical formulation of channel capacity \cite{shannon48,Gallager68,cover06},
the optimization problem formulated above contains some additional
elements\textemdash namely, the decoding metric and the DSP\textemdash which
might be practically useful, though formally redundant. In fact, the
optimal decoding metric is known to be $q(\boldsymbol{y}^{N}|\boldsymbol{x}^{N})=p(\boldsymbol{y}^{N}|\boldsymbol{x}^{N})$.
However, when the true distribution $p(\boldsymbol{y}^{N}|\boldsymbol{x}^{N})$
is either unavailable or too complicated (as in the case of Fig.~\ref{fig:System-description}),
the detector is designed to make maximum-a-posteriori-probability
decisions based on a simpler but mismatched channel law $q(\boldsymbol{y}^{N}|\boldsymbol{x}^{N})\neq p(\boldsymbol{y}^{N}|\boldsymbol{x}^{N})$
\cite{MeKaLaSh94}, which is then optimized subject to some reasonable
assumptions and complexity constraints. The derivation of improved
decoding metrics for the system in Fig.~\ref{fig:System-description}
is discussed, for instance, in \cite{secondini:ecoc17,secondini2019JLT,garcia2020mismatched1,garciagomez2021mismatched}.
Some of these metrics will be used to obtain the capacity lower bounds
shown in Section~\ref{subsec:Combination-of-shaping}.

For what concerns the DSP, its optimization is clearly unnecessary,
since any DSP can be formally included in a proper definition of $p(\boldsymbol{x}^{N})$
and $q(\boldsymbol{y}^{N}|\boldsymbol{x}^{N})$, whose optimization
therefore includes the DSP optimization. However, a properly designed
DSP can simplify the channel $p(\boldsymbol{y}^{N}|\boldsymbol{x}^{N})$
(e.g., by shortening its memory or removing some undesired effects)
and the search for the corresponding optimal input distribution and
decoding metric. Two practical examples that are considered in this
work are chromatic dispersion compensation, which completely removes
the effect of chromatic dispersion in the linear regime, and single-channel
digital backpropagation, which removes deterministic intrachannel
NLI, leaving interchannel NLI as a dominant impairment \cite{Essiambre:JLT0210}.

\section{Input Optimization via Sequence Selection\label{sec:sequence-selection}}

\subsection{General Idea\label{subsec:General-idea}}

\begin{figure}
\begin{centering}
\includegraphics[width=1\columnwidth]{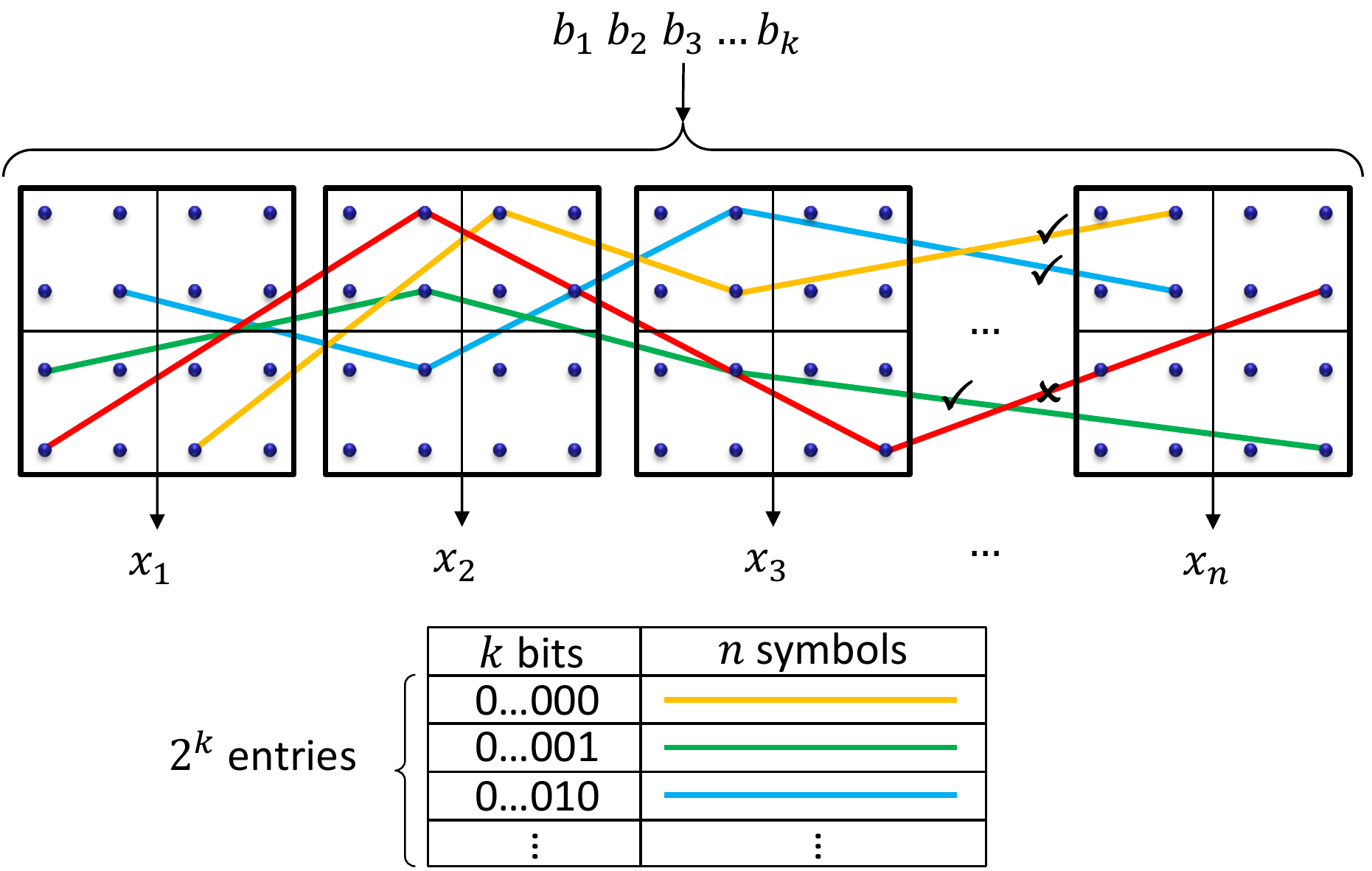}
\par\end{centering}
\caption{\label{fig:sequence_selection_idea}The sequence selection approach:
$k$ input bits are mapped to the ``best'' $2^{k}$ sequences of
$n$ $M$-ary symbols, with rate $R=k/n\le\log_{2}M$.}
\end{figure}

We first illustrate the basic idea with the simple example depicted
in Fig.~\ref{fig:sequence_selection_idea}. Given a discrete-time
channel, we want to find the most efficient way to encode a message
of $k$ information bits on a sequence of $n$ symbols, where each
symbol belongs to a given $M$-ary modulation alphabet (e.g., a 16-QAM
symbol in the figure). Clearly, the problem has no solution if $R\triangleq k/n>\log_{2}M$,
where $R$ is the code rate, as in this case the number of possible
messages $2^{k}$ exceeds the number of available sequences $M^{n}$.
On the other hand, when $R=\log_{2}M$, there is only one trivial
solution, which uses the whole set $\mathcal{A}$ of available sequences
(the particular mapping order being irrelevant). The problem becomes
more interesting when $R<\log_{2}M$. In this case, there are more
sequences than messages, meaning that we can find an optimum map that
uses only the ``best'' $2^{k}$ sequences, contained in a subset
$\mathcal{B\subset A}$. This selection process entails the existence
of a certain \emph{cost function}, which can be used to rank the sequences
according to their cost and to select the ``least expensive'' ones,
discarding the others (e.g., the red one in Fig.~\ref{fig:sequence_selection_idea}).
At this point, the most efficient encoding strategy is simply obtained
by mapping each possible message to a different sequence of the subset
$\mathcal{B}$ (in any arbitrary order), as shown in the table in
Fig.~\ref{fig:sequence_selection_idea}. In practice, by changing
the cardinality of the subset $|\mathcal{B}|=2^{k}$, we can obtain
a different trade-off between code rate $R$ and average cost. For
instance, by selecting less sequences, we reduce the transmission
rate but also the average cost (which means, for instance, a higher
energy-efficiency or a better performance, depending on the considered
cost function).

\subsection{Rejection Sampling Algorithm\label{subsec:Rejection-sampling-algorithm}}

The simple idea illustrated above raises several important issues.
The first issue is related to the procedure used to define the subset
$\mathcal{B\subseteq A}$ of ``best'' sequences. In fact, the proposed
exhaustive search becomes clearly unfeasible for large $M$ and $n$,
or even impossible when considering a continuous input constellation.
Here we propose a different approach, introducing a mechanism to implement
an \emph{optimized source}, i.e., a source that emits random sequences
with an optimized input distribution. The optimization is based on
the same idea presented in Section~\ref{subsec:General-idea}, but
it avoids an exhaustive search and works even in the general case
of a continuous sampling space with large dimensionality. 

\begin{figure}
\begin{centering}
\includegraphics[width=1\columnwidth]{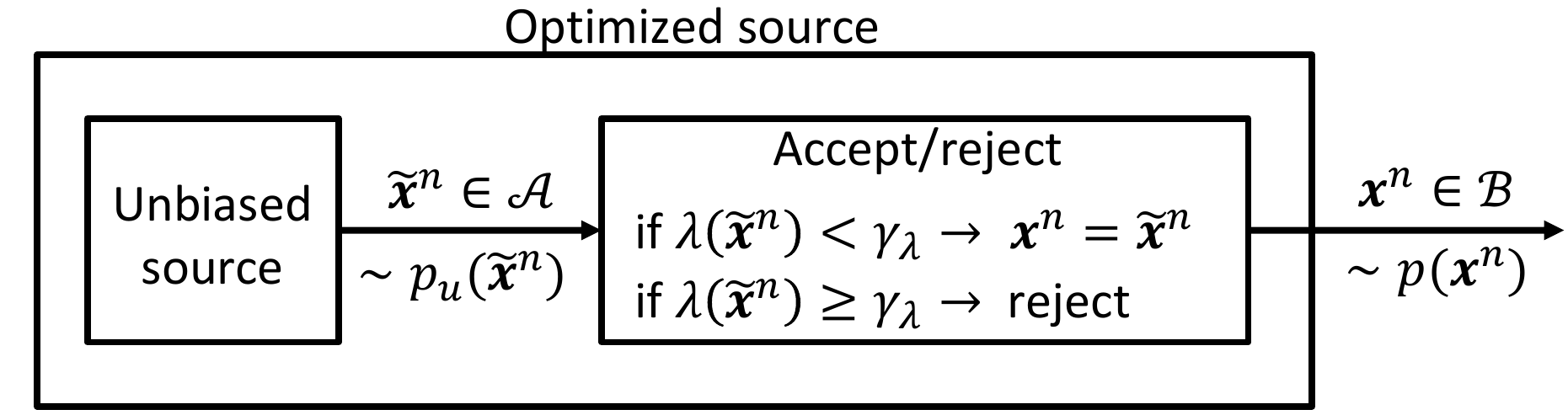}
\par\end{centering}
\caption{\label{fig:rejection_sampling}The rejection sampling machine that
generates sequences with the optimized distribution.}
\end{figure}

The optimized source is implemented by combining an unbiased source
with an accept\textendash reject algorithm, as shown in Fig.~\ref{fig:rejection_sampling}.
 The unbiased source draws a random sequence $\tilde{\boldsymbol{x}}^{n}$
from a certain sampling space $\mathcal{A\subseteq\mathbb{C}}^{n}$
with proposal (unbiased) distribution $p_{u}(\tilde{\boldsymbol{x}}^{n})$.
In general, both the sampling space and the proposal distribution
can be arbitrarily selected, meaning that the symbols can be drawn
from a discrete or continuous constellation, be i.i.d. or not, and
have any distribution. In practice, it is convenient to select a sampling
space that matches the hardware constraints that we have at the transmitter
(e.g., the whole $\mathbb{C}^{n}$ for a theoretical analysis, the
Cartesian product of $n$ QAM constellations for a practical transmitter)
and a proposal distribution that is simple enough in terms of sample
generation and AIR computation, but not too far from the optimal one.
In this work, we will consider sequences of $n$ i.i.d. complex symbols
with a circularly-symmetric Gaussian distribution.

The accept\textendash reject algorithm is characterized by a cost
(or loss) function $\lambda(\tilde{\boldsymbol{x}}^{n})$ and a selection
threshold $\gamma_{\lambda}$. The proposed sequence $\tilde{\boldsymbol{x}}^{n}$
is accepted (emitted by the optimized source by setting $\boldsymbol{x}^{n}=\tilde{\boldsymbol{x}}^{n}$)
if $\lambda(\tilde{\boldsymbol{x}}^{n})<\gamma_{\lambda}$, rejected
otherwise. As a result, the distribution of the sequences emitted
by the optimized source is
\begin{equation}
p(\boldsymbol{x}^{n})=\begin{cases}
\frac{1}{\eta}p_{u}(\boldsymbol{x}^{n}), & \lambda(\boldsymbol{x}^{n})<\gamma_{\lambda}\\
0, & \lambda(\boldsymbol{x}^{n})\geq\gamma_{\lambda}
\end{cases}\label{eq:prob_biased}
\end{equation}
where
\begin{equation}
\eta=\mathrm{Pr}\{\lambda(\tilde{\boldsymbol{X}}^{n})<\gamma_{\lambda}\}\label{eq:acceptance-rate-definition}
\end{equation}
is the acceptance probability, which can be practically estimated
as $\eta\approx N_{a}/N_{p}$ , where $N_{a}$ is the number of accepted
sequences and $N_{p}$ is the number of proposed sequences. On average,
each accepted sequence requires the generation of $1/\eta$ proposed
sequences and the computation of the corresponding $1/\eta$ values
of the cost function. This gives an indication of the computational
cost of the proposed procedure.

In practice, the algorithm guarantees that all the emitted sequences
belong to the set $\mathcal{B\subseteq A}$ of ``best'' sequences
(whose cost is lower than the prescribed threshold $\gamma_{\lambda}$);
within such set, their distribution is just a rescaled version of
the unbiased one. Reducing (increasing) $\gamma_{\lambda}$ reduces
the average cost of the sequences emitted by the optimized source.
At the same time, it reduces also the acceptance rate $\eta$, which
in turn reduces the (differential) entropy rate of the optimized source
(in the example of Fig.~\ref{fig:sequence_selection_idea}, the number
of sequences $|\mathcal{B}|=2^{k}$ available to encode information
decreases, reducing the rate $k/n$) and increases the computational
complexity of the generation procedure.

\subsection{AIR Estimation\label{subsec:AIR-estimation}}

In principle, the AIR with mismatched decoding and the optimized source
described in Section~\ref{subsec:Rejection-sampling-algorithm} can
be estimated by the general expression in (\ref{eq:AIR-general}),
where the input sequence $\boldsymbol{x}^{N}$ is obtained by concatenating
several independent subsequences of length $n$ emitted by the optimized
source (for simplicity, we assume that $N/n$ is an integer number).
The corresponding input distribution $p(\boldsymbol{x}^{N})$ is hence
the product of the distributions of the $N/n$ subsequences, given
by (\ref{eq:prob_biased}), while the mismatched output distribution
$q(\boldsymbol{y}^{N})$ is given by (\ref{eq:mismatched_output_distribution}).
However, since (\ref{eq:prob_biased}) is a complicated distribution
in an $n$-dimensional space, the computation of (\ref{eq:mismatched_output_distribution})
can be very complex in this case. Thus, we resort to the lower bound
\begin{multline}
I_{\eta}(\boldsymbol{X};\boldsymbol{Y})=\lim_{N\rightarrow\infty}\frac{1}{N}E\left\{ \log_{2}\frac{q(\boldsymbol{Y}^{N}|\boldsymbol{X}^{N})}{q_{u}(\boldsymbol{Y}^{N})}\right\} \\
-\frac{1}{n}\log_{2}\frac{1}{\eta}\le I_{q}(\boldsymbol{X};\boldsymbol{Y})\label{eq:AIR_rate_loss}
\end{multline}
where the expectation is taken with respect to the actual optimized
distribution; the inequality is obtained by using $p(\boldsymbol{x}^{N})\leq p_{u}(\boldsymbol{x}^{N})/\eta^{N/n}$,
which follows from (\ref{eq:prob_biased}); and
\begin{equation}
q_{u}(\boldsymbol{y}^{N})=\int q(\boldsymbol{y}^{N}|\boldsymbol{x}^{N})p_{u}(\boldsymbol{x}^{N})d\boldsymbol{x}^{N}\label{eq:unbiased_output_distribution}
\end{equation}
is obtained by connecting the unbiased source $p_{u}(\boldsymbol{x}^{N})$
to the auxiliary channel $q(\boldsymbol{y}^{N}|\boldsymbol{x}^{N})$
and is, hence, much simpler to compute than $q(\boldsymbol{y}^{N})$.
For instance, for an unbiased source of i.i.d. Gaussian samples with
variance $P$ and an AWGN decoding metric with variance $\sigma^{2}$,
the samples at the output of the auxiliary channel are i.i.d. Gaussian
with variance $P+\sigma^{2}$. In practice, the AIR (\ref{eq:AIR_rate_loss})
can be estimated by the following simple procedure:
\begin{itemize}
\item create a long input sequence $\boldsymbol{x}^{N}$ by concatenating
$N/n$ subsequences of length $n$ generated by the rejection sampling
machine in Fig.~\ref{fig:rejection_sampling};
\item compute numerically (or generate experimentally) the corresponding
output sequence $\boldsymbol{y}^{N}$ obtained by the propagation
of $\boldsymbol{x}^{N}$ through the system in Fig.~\ref{fig:System-description};
\item estimate the AIR using
\begin{equation}
I_{\eta}(X;Y)\approx\frac{1}{N}\log_{2}\frac{q(\boldsymbol{y}^{N}|\boldsymbol{x}^{N})}{q_{u}(\boldsymbol{y}^{N})}-\frac{1}{n}\log_{2}\frac{N_{p}}{N_{a}}
\end{equation}
i.e., using the same expression as in the unbiased case and subtracting
the rate loss.
\end{itemize}
The AIR can then be maximized by optimizing the selection threshold
$\gamma_{\lambda}$, i.e., the acceptance rate $\eta\approx N_{a}/N_{p}$.
The overall procedure is very similar to the case without sequence
selection, the only additional complexity being the generation of
the input symbols based on the rejection sampling machine in Fig.~\ref{subsec:Rejection-sampling-algorithm},
which requires $1/\eta$ computations of the cost function for each
accepted subsequence of length $n$.

\subsection{Cost Function}

In practice, the definition of a suitable cost function is essential
for the correct optimization of the source. A good cost function must
be simple enough to be computed several times per each generated sequence
and, at the same time, accurate enough to yield an effective selection.
The final goal is that of maximizing the AIR (\ref{eq:AIR-general})
over the given channel, subject to some possible constraints. This
means that, in general, the cost function should be specifically defined
to account for the channel characteristics, the decoding metric, and
the required constraints.

As a first example, we consider a simple AWGN channel with matched
detection and an average-power constraint. In this case, the channel
effect is independent of the particular input sequence and depends
only on the noise variance. On the other hand, different sequences
may have a different energy, meaning that their use may have a different
cost in terms of average power. Therefore, given a generic input sequence
$\boldsymbol{s}^{n}$, a suitable cost function appears to be the
\emph{energy per symbol} of the sequence
\begin{equation}
\lambda(\boldsymbol{s}^{n})=\frac{1}{n}\left\Vert \boldsymbol{s}^{n}\right\Vert ^{2}\label{eq:cost_function_energy}
\end{equation}
With reference to the example in Fig.~\ref{fig:sequence_selection_idea},
setting a particular threshold $\gamma_{\lambda}$ for the cost function
in (\ref{eq:cost_function_energy}) corresponds to select all the
sequences within the sphere of radius $\sqrt{n\gamma_{\lambda}}$
in $\mathbb{C}^{n}$. Changing the threshold $\gamma_{\lambda}$ changes
the number of selected sequences $2^{k}$, i.e., the rate $R=k/n$
of the shaping strategy. Among all possible strategies, the sphere
shaping induced by the cost function (\ref{eq:cost_function_energy})
clearly minimizes the average power required to achieve the desired
rate.

The situation is different for the nonlinear channel, which induces
also signal-dependent distortions due to fiber nonlinearity. At the
same time, the average power constraint might be no longer relevant
in this case, since the AIR has typically a peaky behavior and the
goal is that of maximizing the peak AIR, regardless of the average
power. The impact of the channel on the AIR (\ref{eq:AIR_rate_loss})
is determined by the (mismatched) conditional entropy term $-E\left\{ \log_{2}q(\boldsymbol{Y}^{N}|\boldsymbol{X}^{N})\right\} /N$,
which measures the average uncertainty that we have about the output
sequence, given the input sequence and the mismatched probabilistic
model $q(\boldsymbol{y}^{N}|\boldsymbol{x}^{N})$.\footnote{Some authors use the term ``cross entropy'' to refer to the quantity
$-E_{p}\{\log q(X)\}$, where $p(x)$ and $q(x)$ are two distributions
defined over the same probability space, and $E_{p}\{\cdot\}$ denotes
expectation with respect to $p$. The same term is used by other authors
to refer to the relative entropy or Kullback\textendash Leibler divergence
$-E_{p}\{\log(q(X)/p(X))$ \cite{cover06}. To avoid confusion and
better highlight the relation with the mismatched probabilistic model,
we prefer here the term ``mismatched entropy''.} Therefore, a suitable cost function to minimize this term (and maximize
the AIR) is
\begin{equation}
\lambda(\boldsymbol{s}^{n})=-\frac{1}{N}E\left\{ \log_{2}q(\boldsymbol{Y}^{N}|\boldsymbol{X}^{N})|\boldsymbol{X}_{N/2-n/2+1}^{N/2+n/2}=\boldsymbol{s}^{n})\right\} \label{eq:cost_function_conditional_entropy}
\end{equation}
which measures the average effect that the particular sequence $\boldsymbol{s}^{n}$
has on the mismatched conditional entropy term. Note that, in the
case of an AWGN channel with matched decoding, the cost function (\ref{eq:cost_function_conditional_entropy})
equals the entropy of the noise and is independent of the sequence
$\boldsymbol{s}^{n}$, meaning that all the sequences are equally
good. In this case, the introduction of an average power constraint
and the use of the cost function (\ref{eq:cost_function_energy})
is clearly more appropriate.

It is instructive to specialize the cost function above to the case
of an AWGN decoding metric 
\begin{equation}
q(\boldsymbol{y}^{N}|\boldsymbol{x}^{N})=\frac{1}{\pi\sigma^{2N}}\exp\left(-\frac{\left\Vert \boldsymbol{y}^{N}-\boldsymbol{x}^{N}\right\Vert ^{2}}{\sigma^{2}}\right)\label{eq:AWGN metric}
\end{equation}
By replacing (\ref{eq:AWGN metric}) in (\ref{eq:cost_function_conditional_entropy})
and omitting some inessential constant terms (which do not alter the
ranking induced by the cost function over the available sequences),
the cost function can be expressed as
\begin{equation}
\lambda(\boldsymbol{s}^{n})=\frac{1}{N}E\left\{ \left\Vert \boldsymbol{Y}^{N}-\boldsymbol{X}^{N}\right\Vert ^{2}|\boldsymbol{X}_{N/2-n/2+1}^{N/2+n/2}=\boldsymbol{s}^{n}\right\} \label{eq:cost-function_AWGNmetric}
\end{equation}
In practice, (\ref{eq:cost-function_AWGNmetric}) can be estimated
by replacing the expectation with an average over a finite number
of realizations, and considering only a shorter portion of the sequence,
i.e., only those symbols which are more affected by the input sequence
$\boldsymbol{s}^{n}$\textemdash the $n$ central symbols and, possibly,
a few surrounding ones, depending on the channel memory.

A final simplification can be obtained by assuming that nonlinear
effects are block-memoryless, with blocklength $n$, and independent
of the noise. In this case, both the noise and the symbols outside
the block of length $n$ become irrelevant, as they contribute only
with a constant term to the cost function (\ref{eq:cost-function_AWGNmetric}),
which can therefore be simplified as
\begin{equation}
\lambda(\boldsymbol{s}^{n})=\frac{1}{n}\left\Vert \hat{\boldsymbol{y}}^{n}-\boldsymbol{s}^{n}\right\Vert ^{2}\label{eq:cost_function_memoryless}
\end{equation}
where the vector $\hat{\boldsymbol{y}}^{n}$ collects the $n$ output
symbols obtained from a noiseless propagation of the input sequence
$\boldsymbol{x}^{n}=\boldsymbol{s}^{n}$. The block-memoryless assumption
ensures that $\hat{\boldsymbol{y}}^{n}$ depends deterministically
on $\boldsymbol{s}^{n}$, so that the expectation is no longer required
and the cost function can be estimated from a single noiseless simulation.

\section{Analytical Results\label{sec:Analytical-results}}

We apply here the proposed technique to a simple nonlinear channel
that includes AWGN and block-memoryless NLI. Given a block of $n$
input symbols $\boldsymbol{x}^{n}$, the corresponding output block
can be written as
\begin{equation}
\boldsymbol{y}^{n}=\boldsymbol{x}^{n}+\boldsymbol{w}^{n}+\boldsymbol{\xi}^{n}\label{eq:block_memoryless_RPchannel}
\end{equation}
where $\boldsymbol{w}^{n}$ is a vector of i.i.d. noise samples with
circularly symmetric complex Gaussian distribution with variance $\sigma_{w}^{2}$,
and $\boldsymbol{\xi}^{n}$ is a vector of NLI samples. As in a regular-perturbation-based
fiber channel model \cite{Mecozzi:JLT0612,Dar2013:opex}, we assume
that NLI samples are generated by the nonlinear interaction of the
input symbols during propagation and scale cubically with them, so
that their variance can be written as $\sigma_{\xi}^{2}=aP^{3}$,
where $P=E\{|X_{i}|^{2}\}$ is the input power and $a$ a proportionality
constant. Moreover, we assume that channel memory does not extend
beyond the edges of each block, i.e., separate blocks are independent.
We remark that (\ref{eq:block_memoryless_RPchannel}), with the related
assumptions, is not intended as an accurate description of a realistic
fiber channel but, rather, as a simplified model that allows for an
analytical study of the problem while retaining some essential features
of optical fiber channels.

\begin{figure}
\begin{centering}
\includegraphics[width=1\columnwidth]{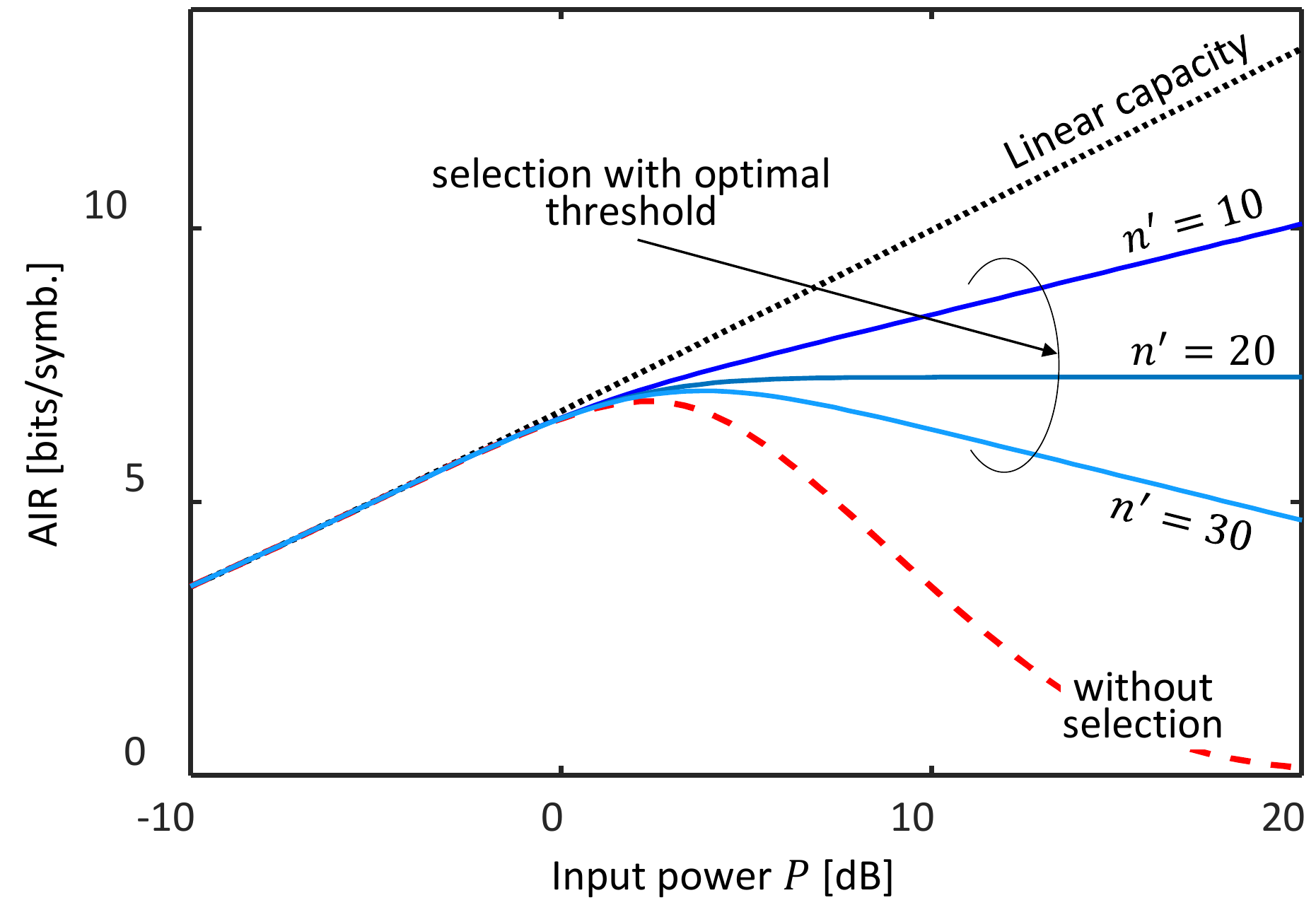}
\par\end{centering}
\caption{\label{fig:AIR_analytic}AIR over the nonlinear block-memoryless channel
with Gaussian decoding metric, with and without sequence selection.}
\end{figure}

First, we consider the simple case in which the system is designed
according to a simple Gaussian noise model, i.e., by completely neglecting
the dependence of NLI on the input symbols and by assuming that the
elements of $\boldsymbol{\xi}^{n}$ are i.i.d. circularly symmetric
complex Gaussian variables \cite{Pog:JLT12}. In this case, (\ref{eq:block_memoryless_RPchannel})
reduces to an AWGN channel with noise variance $\sigma^{2}=\sigma_{w}^{2}+\sigma_{\xi}^{2}$,
for which the optimal input distribution consists in i.i.d. circularly
symmetric complex Gaussian symbols, and the optimal decoding metric
is given by (\ref{eq:AWGN metric}). The corresponding AIR is
\begin{equation}
I_{G}(X;Y)=\log_{2}\left(1+\frac{P}{\sigma_{w}^{2}+aP^{3}}\right),\label{eq:AIR_gaussian}
\end{equation}
which has the typical behavior shown in Fig.~\ref{fig:AIR_analytic}
with a red dashed line for the case $a=0.01$ and $\sigma_{w}^{2}=0.001$.
With respect to the linear capacity $C=\log_{2}(1+P/\sigma_{w}^{2})$
obtained for $a=0$, which grows unbounded with power, the Gaussian
AIR (\ref{eq:AIR_gaussian}) over the nonlinear channel reaches a
peak at the optimal input power $P_{\mathrm{opt}}=\sqrt[3]{\sigma_{w}^{2}/2a}$,
after which it decreases again and vanishes. Though this behavior
appears often in the literature with reference to the optical fiber
channel or its approximated models, it has been already pointed out
that there is no proof that a similar behavior applies to the actual
channel capacity. Indeed, a different behavior has been demonstrated
for different approximated channel models, whereas the problem remains
open when considering a realistic channel \cite{secondini_JLT2017_scope,SECONDINI2020867}.

Then, we consider a different approach, in which we account for the
dependence of NLI on the input symbols in (\ref{eq:block_memoryless_RPchannel})
to better shape the input distribution. In particular, we optimize
the distribution of the input symbols by applying the sequence selection
procedure described in Section~\ref{sec:sequence-selection} to an
unbiased source of i.i.d. Gaussian symbols with power $P$. The length
of the input sequences is taken equal to the block length $n$ over
which the channel memory extends. On the other hand, we do not optimize
the decoding metric, keeping the simple AWGN metric in (\ref{eq:AWGN metric}).
For this metric, given the block-memoryless property of the channel,
a suitable cost function is the energy per symbol of the NLI (\ref{eq:cost_function_memoryless}),
which in this case can be expressed as
\begin{equation}
\lambda(\boldsymbol{x}^{n})=\frac{1}{n}\left\Vert \xi^{n}(\boldsymbol{x}^{n})\right\Vert ^{2}\label{eq:cost_function_analytic-AIR}
\end{equation}
 with expectation $E\{\Lambda\}=aP^{3}$. The AIR obtained with the
optimized source is given by (\ref{eq:AIR_rate_loss}) and can be
expressed as 
\begin{equation}
I_{\eta}(\boldsymbol{X};\boldsymbol{Y})=\log_{2}\left(1+\frac{P}{\sigma_{n}^{2}+\sigma_{\xi}^{2}(\gamma_{\lambda})}\right)-\frac{1}{n}\log_{2}\frac{1}{\eta(\gamma_{\lambda})}\label{eq:AIR_selection}
\end{equation}
where we have made explicit the dependence on the threshold $\gamma_{\lambda}$
of both the acceptance rate $\eta(\gamma_{\lambda})$ and the NLI
variance (after selection) $\sigma_{\xi}^{2}(\gamma_{\lambda})$,
and we have assumed that $E\{|Y|^{2}\}=P+\sigma_{n}^{2}+\sigma_{\xi}^{2}(\gamma_{\lambda})$.\footnote{In practice, we assume that the output power equals the input power
plus the noise and NLI power, implying that the three processes are
uncorrelated. This is clearly a simplification, since NLI does depend
on the input symbols. Nonetheless, the following analysis would produce
qualitatively similar results by making more realistic assumptions,
e.g., $E\{|Y|^{2}\}=P+\sigma_{n}^{2}$ (the total energy is preserved
by the nonlinear propagation), or $E\{|Y|^{2}\}=P+\sigma_{n}^{2}-\sigma_{\xi}^{2}(\gamma_{\lambda})$
(the NLI is anticorrelated with the signal and reduces the output
power, for instance due to the loss induced by spectral broadening).} By reducing $\gamma_{\lambda}$, we accept only sequences with a
lower NLI, reducing $\sigma_{\xi}^{2}(\gamma_{\lambda})$. At the
same time, we accept less sequences, reducing also $\eta(\gamma_{\lambda})$.
In terms of AIR, the first effect is beneficial, whereas the second
one is detrimental. The opposite behavior is obtained by increasing
$\gamma_{\lambda}$. Two special cases are obtained for $\gamma_{\lambda}\rightarrow0$
and $\gamma_{\lambda}\rightarrow\infty$. In the first case, $\eta\rightarrow0$
and the information rate vanishes (no sequences are available). In
the second case, $\eta\rightarrow1$, the source remains unbiased,
and $\sigma_{\xi}^{2}(\gamma_{\lambda})\rightarrow aP^{3}$, so that
(\ref{eq:AIR_selection}) reduces to (\ref{eq:AIR_gaussian}).

In order to study the more general case and find the optimal threshold
$\gamma_{\lambda}$ that maximizes the AIR, we need to make some additional
assumptions about the NLI distribution. In particular, we assume that
when the input is unbiased (i.i.d. Gaussian symbols), the cost function
in (\ref{eq:cost_function_analytic-AIR}) has a gamma distribution
with shape parameter $n'\le n$ and expectation $E\{\Lambda\}=aP^{3}$
\begin{equation}
p_{\lambda}(\lambda)=\frac{1}{\Gamma(n')}\left(\frac{n'}{aP^{3}}\right)^{n'}\lambda^{n'-1}\exp\left(-\frac{n'\lambda}{aP^{3}}\right),\quad\lambda>0\label{eq:distribution_cost_function}
\end{equation}
where $\Gamma(n)$ is the gamma function \cite[eq. (6.1.1)]{Abramowitz:1972}.
This assumption is motivated by the observation that, if NLI were
indeed statistically equivalent to AWGN, as suggested by the GN model,
the distribution of the cost function would be exactly the one in
(\ref{eq:distribution_cost_function}) but with shape $n'=n$\textemdash i.e.,
a rescaled chi-squared distribution with $2n$ degrees of freedom.
On the other hand, adjacent NLI samples are not really independent,
as the same input symbols are involved in the generation of several
consecutive NLI samples, so that it is reasonable to assume that the
actual shape parameter in (\ref{eq:distribution_cost_function}) might
be reduced. Again, (\ref{eq:distribution_cost_function}) is just
a working assumption, with no claims of accuracy. In the next section,
more realistic fiber models will be studied numerically, also showing
how far the assumptions of this section are from reality.

The dependence of the acceptance rate and variance on the threshold
can be computed from the distribution in (\ref{eq:distribution_cost_function}).
The acceptance rate equals the cumulative distribution function (cdf)
\begin{equation}
\eta(\gamma_{\lambda})=\int_{0}^{\gamma_{\lambda}}p_{\lambda}(\lambda)d\lambda=\frac{\gamma\left(n',n'\gamma_{\lambda}P^{-3}/a\right)}{\Gamma(n')}\label{eq:acceptance_rate_gamma}
\end{equation}
where $\gamma(s,x)$ is the lower incomplete gamma function with parameter
$s$ \cite[eq. (6.5.2)]{Abramowitz:1972}. The NLI variance equals
the conditional expectation 
\begin{align}
\sigma_{\xi}^{2}(\gamma_{\lambda}) & =E\{\Lambda|\Lambda<\gamma_{\lambda}\}=\int_{0}^{\gamma_{e}}\lambda\frac{p_{\lambda}(\lambda)}{\eta(\gamma_{\lambda})}d\lambda\nonumber \\
 & =\frac{aP^{3}}{n'}\frac{\gamma\left(n'+1,n'\gamma_{\lambda}P^{-3}/a\right)}{\gamma\left(n',n'\gamma_{\lambda}P^{-3}/a\right)}\label{eq:nli_variance_gamma}
\end{align}
The AIR is eventually obtained by replacing (\ref{eq:acceptance_rate_gamma})
and (\ref{eq:nli_variance_gamma}) in (\ref{eq:AIR_selection})
\begin{align}
I_{\eta}(\boldsymbol{X};\boldsymbol{Y}) & =\log_{2}\left(1+\frac{P}{\sigma_{w}^{2}+\frac{aP^{3}}{n'}\frac{\gamma\left(n'+1,n'\gamma_{\lambda}P^{-3}/a\right)}{\gamma\left(n',n'\gamma_{\lambda}P^{-3}/a\right)}}\right)\nonumber \\
 & \quad-\frac{1}{n}\log_{2}\frac{\Gamma(n')}{\gamma\left(n',n'\gamma_{\lambda}P^{-3}/a\right)}\label{eq:AIR_analytic}
\end{align}

Interestingly, while the Gaussian AIR (\ref{eq:AIR_gaussian}) has
the typical peaky behavior that is commonly found for most available
capacity lower bounds for the nonlinear optical fiber\textemdash it
reaches a peak at some optimal power and then decays again to zero
at high power\textemdash this is not necessarily the case for the
AIR (\ref{eq:AIR_analytic}) with sequence selection. In fact, by
setting the threshold to the approximately optimal value 
\begin{equation}
\gamma_{\lambda}^{\mathrm{opt}}\approx\frac{n'+1}{n-n'}\sigma_{n}^{2}\label{eq:optimal_threshold}
\end{equation}
three different asymptotic behaviors can be observed for $P\rightarrow\infty$:
i) for $n'<n/3$, the AIR grows unbounded as $\log_{2}(P^{1-3n'/n})$;
ii) for $n'=n/3$, the AIR saturates to a finite constant value; for
$n'>n/3$, the AIR vanishes after reaching a finite peak at some optimum
power (as in the unshaped case). This result, obtained specifically
for the distribution in (\ref{eq:distribution_cost_function}), is
more general and holds whenever the cdf in (\ref{eq:acceptance_rate_gamma})
grows as $\sim\gamma_{\lambda}^{n'}$ for small $\gamma_{\lambda}$.
A slow growth ($n'<n/3$) implies that the AIR can always be increased
by increasing the power $P$ and simultaneously decreasing the selection
threshold $\gamma_{\lambda}$ to keep the NLI limited, since the rate
loss caused by the selection is more than compensated by the corresponding
increase of the SNR. On the other hand, this is not possible, at least
asymptotically, when the growth is faster ($n'>n/3$).

Fig.~\ref{fig:AIR_analytic} shows the AIR (\ref{eq:AIR_selection})
with optimized selection threshold for the block-memoryless nonlinear
channel (\ref{eq:block_memoryless_RPchannel}) with nonlinear coefficient
$a=0.01$, noise variance $\sigma_{w}^{2}=0.001$, block length $n=60$,
under the assumption that the NLI intensity, averaged over the block,
has the distribution in (\ref{eq:distribution_cost_function}) with
different values of the scale parameter $n'$. For comparison, the
figure reports also the linear channel capacity and the Gaussian AIR
(\ref{eq:AIR_gaussian}). As anticipated, three fundamentally different
asymptotic behaviors (unbounded growth, saturation, or decay to zero)
are observed, depending on $n'$. Even in the worst case ($n'=30>n/3$),
in which the AIR decays to zero for $P\rightarrow\infty$, a gain
with respect to the unshaped case is obtained in the nonlinear regime.

\section{Numerical Results\label{sec:Numerical-results}}

\subsection{System Description\label{subsec:System-description}}

The system is depicted in Fig.\ref{fig:System-description}, while
the scenario and link parameters are the same considered in \cite{secondini2019JLT,garciagomez2021mismatched}\textemdash a
1000~km standard single-mode fiber link with attenuation $\alpha=\unit[0.2]{dB/km}$,
group-velocity-dispersion parameter $\beta_{2}=\unit[21.7]{ps^{2}/km}$,
and nonlinear coefficient $\gamma=\unit[1.27]{W^{-1}km^{-1}}$; ideal
distributed amplification with unitary spontaneous emission coefficient;
and, unless otherwise specified, five dual-polarization 50~GBd Nyquist-WDM
channels with sinc pulse shape and 50~GHz spacing.\footnote{The same scenario was considered also in \cite{Secondini:2021ECOC},
though we accidentally used a slightly higher nonlinear coefficient
$\gamma=\unit[1.3]{W^{-1}km^{-1}}$, which explains why the AIR values
in \cite{Secondini:2021ECOC} are slightly lower in the nonlinear
regime.} The transmitted symbols are i.i.d. circularly symmetric Gaussian
variables (unbiased distribution), further processed by the rejection
sampling machine in Fig.~\ref{fig:rejection_sampling} when sequence
selection is applied (optimized distribution). At the receiver, the
central WDM channel is demultiplexed by using an ideal rectangular
filter with 50~GHz bandwidth and further processed according to the
selected DSP (ideal dispersion compensation or ideal single-channel
DBP) and decoding metric (optimized for the AWGN or PPN channel).
When subcarrier multiplexing is considered, each WDM channel is divided
into four 12.5~GBd subcarriers with sinc pulse shape and 12.5~GHz
spacing. Before detection, a possible constant average phase rotation
affecting the signal (due to nonlinearity) is estimated and removed.

Fiber propagation is emulated by using the split step Fourier method
(SSFM); a step size of $\unit[100]{m}$ and a sampling rate of 400~GHz
(eight samples per symbol in the single-carrier case) are used for
the WDM scenario; a step size of $\unit[500]{m}$ and a sampling rate
of 100~GHz (two samples per symbol in the single-carrier case) for
the single-channel scenario, for the computation of the cost function,
and for the implementation of DBP.

The results of this section are shown in terms of achievable SE, measured
in bits/s/Hz/pol and estimated by the procedure described in Section~\ref{subsec:AIR-estimation}
with $N=2^{18}$ symbols. As a benchmark, all the figures report also
the SE obtained on the nonlinear channel when the system is optimized
in the absence of nonlinear effects, i.e., for a system with ideal
dispersion compensation, i.i.d. Gaussian input symbols, and AWGN detection.
The capacity per unit bandwidth for the linear channel $C=\log_{2}(1+\mathrm{SNR})$
is also reported as a reference.

\subsection{Single-Channel Single-Polarization System\label{subsec:Single-channel-single-polarizati}}

The first numerical test is performed in a simple scenario, considering
a single-polarization single-channel system with ideal dispersion
compensation. The main aim of this test is to study the behavior of
the proposed strategy in a more realistic channel, where intrachannel
NLI (including signal-noise interaction and spectral broadening) is
accurately modeled by the SSFM, possibly departing from the assumptions
made for the analytic study in Section~\ref{sec:Analytical-results}
(block-memoryless, cubic scaling with power, gamma distribution).
Moreover, the test will also show how far those assumptions are from
reality, and how fast the cdf of the NLI energy grows with respect
to the critical rate defined in Section~\ref{sec:Analytical-results}.

In this scenario, we employ the simple cost function in (\ref{eq:cost_function_memoryless}),
without averaging over the realizations. In practice, the selection
is implemented by the following \emph{fast} procedure:
\begin{itemize}
\item generate a very long input sequence $\boldsymbol{x}^{N'}$ by drawing
samples from the unbiased distribution and evaluate the corresponding
output sequence $\hat{\boldsymbol{y}}^{N'}$ by SSFM propagation through
the noiseless channel;
\item take all the $N_{p}=N'-n+1$ subsequences of length $n$, $\{\boldsymbol{x}_{i+1}^{i+n}\}$,
for $i=0,\ldots,N_{p}-1$, that are contained in $\boldsymbol{x}^{N'}$
as proposed sequences;
\item for each proposed sequence, compute the cost function 
\begin{equation}
\lambda(\boldsymbol{x}_{i+1}^{i+n})=\left\Vert \hat{\boldsymbol{y}}_{i+1}^{i+n}-\boldsymbol{x}_{i+1}^{i+n}\right\Vert ^{2}
\end{equation}
\item accept the $N_{a}$ subsequences for which the cost function is below
threshold.
\end{itemize}
If more sequences are needed, the procedure is repeated several times.
The selection is performed at a single launch power (near the optimum
launch power without selection), under the hypothesis that the ranking
induced by the cost function on the sequences does not change significantly
with power. This is true, for instance, if the NLI scales cubically
with power. The AIR and SE at the desired launch power are eventually
estimated by using the procedure described in Section~\ref{subsec:AIR-estimation},
where the input sequence $\boldsymbol{x}^{N}$ is formed by concatenating
the accepted sequences.

\begin{figure*}
\begin{centering}
\includegraphics[width=1\columnwidth]{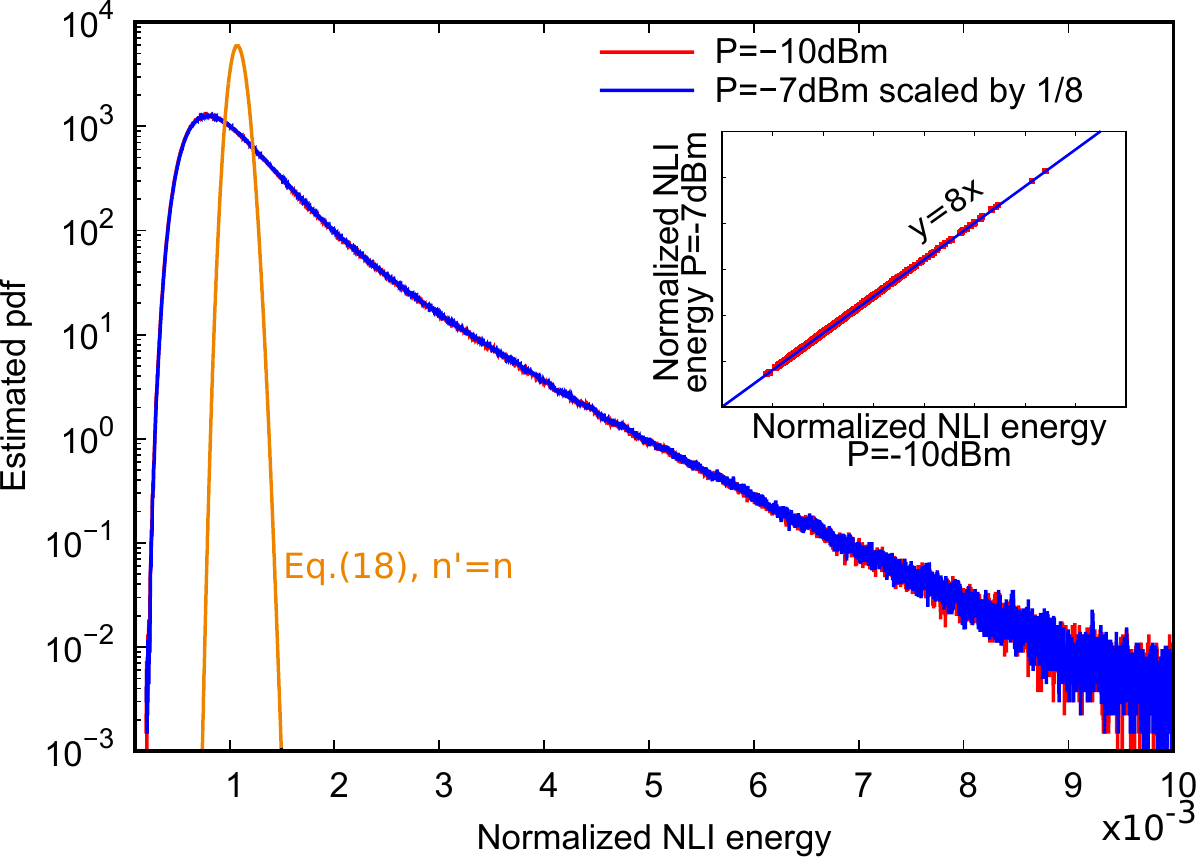}\includegraphics[width=1\columnwidth]{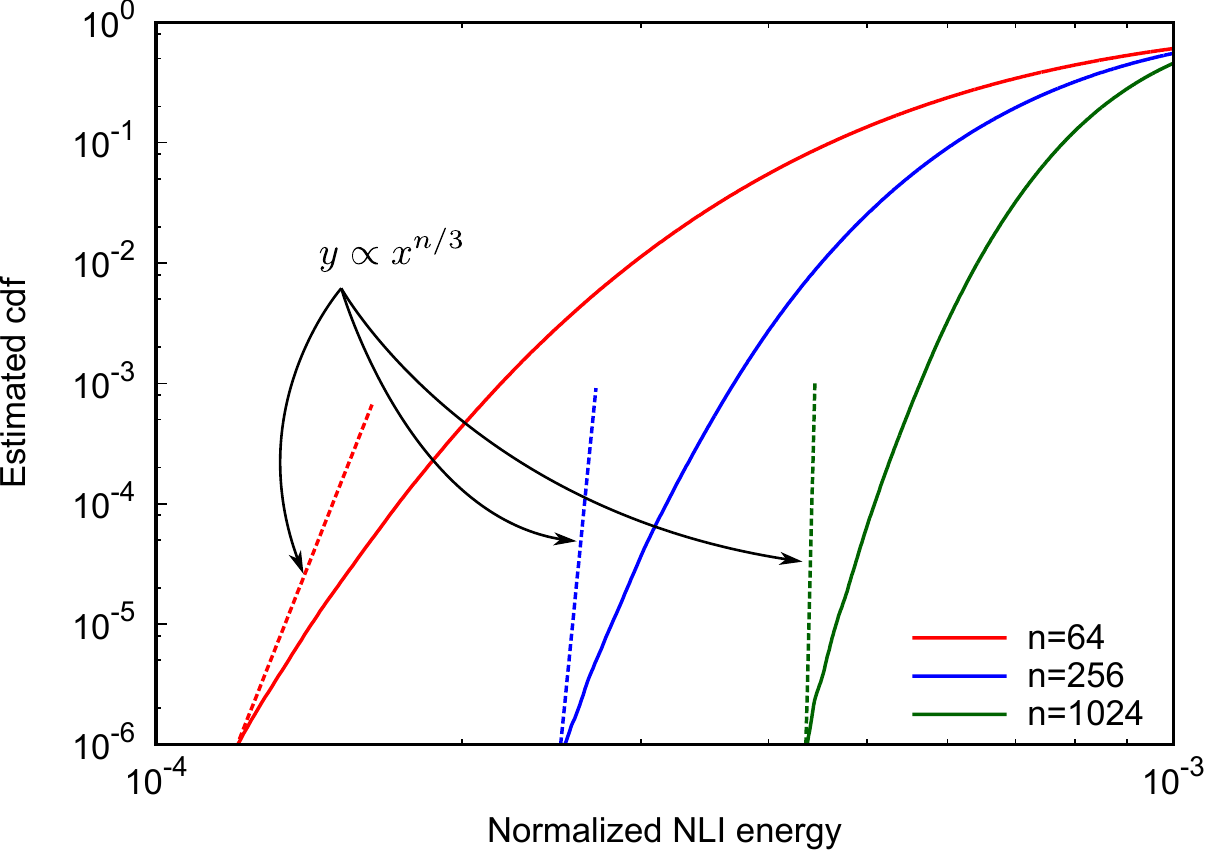}
\par\end{centering}
\hspace*{\fill}(a)\hspace*{\fill}\hspace*{\fill}(b)\hspace*{\fill}

\caption{\label{fig:pdf_cdf}Distribution of the normalized NLI energy in the
single-channel single-polarization system: (a) empirical distribution
for $n=256$ and two different powers (the coordinates of the points
in the inset correspond to the realizations obtained for each sequence
and the two power values) compared to the gamma distribution (\ref{eq:distribution_cost_function})
with $n'=n$; (b) empirical cdf for different blocklength $n$ compared
to the power law $\sim\lambda^{n/3}$ needed for an unbounded growth
of the AIR.}
\end{figure*}

First, we compare the assumptions made in Section~\ref{sec:Analytical-results}
about the simplified nonlinear channel (\ref{eq:block_memoryless_RPchannel})
with the actual NLI measured in the more realistic channel considered
here. Fig.~\ref{fig:pdf_cdf}(a) shows the distribution of the cost
function (\ref{eq:cost_function_memoryless}), corresponding to the
NLI energy per symbol, evaluated for $N_{p}\approx1.35\times10^{9}$
proposed sequences with blocklength $n=256$ and normalized to the
mean signal energy per symbol. The empirical distributions obtained
at two different powers (one rescaled by the cube of the power ratio
to match the other) are compared to the gamma distribution (\ref{eq:distribution_cost_function})
with $n'=n$ that would be obtained for the energy of an AWGN process.
The two empirical distributions are practically superimposed, confirming
the assumption that the NLI scales with very good approximation with
the cube of the input power. This is even better illustrated by the
inset of Fig.~\ref{fig:pdf_cdf}(a), where a point is reported for
each tested sequence, with coordinates equal to the NLI obtained for
the two different launch powers. As expected, all the points practically
lie on the straight line $y=8x$. Moreover, the empirical distributions
of the NLI energy are significantly different and wider than the distribution
that would be obtained if NLI were statistically equivalent to an
AWGN, meaning that the AWGN assumption typically made in GN models
is too pessimistic and not accurate enough for this approach. This
is even better appreciated in Fig.~\ref{fig:pdf_cdf}(b), which reports
the left tails of the empirical cdf obtained for different values
of the block length $n$. The results show that the particular characteristics
of the generated NLI and the dependence between its samples make its
cdf grow with an exponent that is not only lower than the value $n$
that would be obtained for AWGN, but also much lower than the highest
exponent $n/3$ that guarantees an unbounded growth of AIR with power
according to the theory developed in Section~\ref{sec:Analytical-results}.
This is clearly not a proof that the same unbounded growth can be
actually obtained on this more realistic channel, but at least a good
reason to expect a significant AIR improvement. 

\begin{figure*}
\begin{centering}
\includegraphics[width=1\columnwidth]{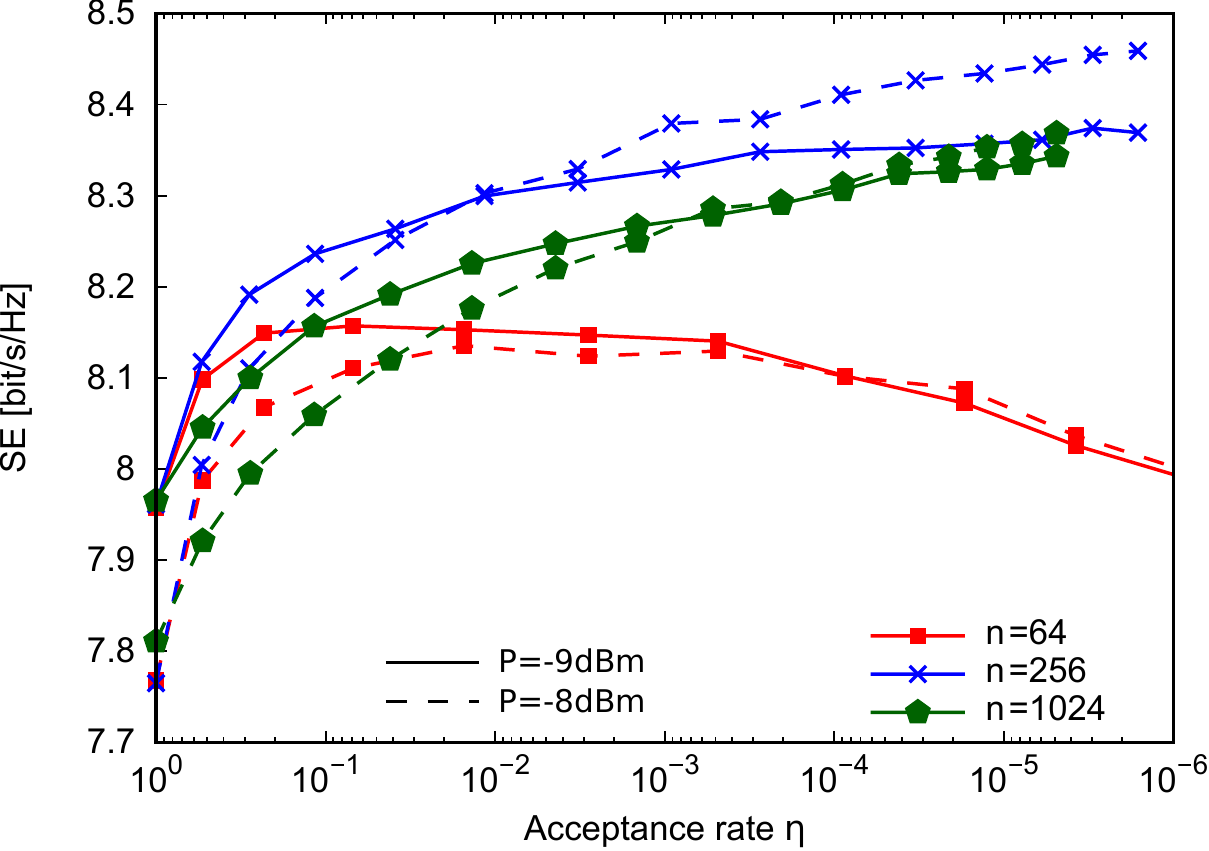}\includegraphics[width=1\columnwidth]{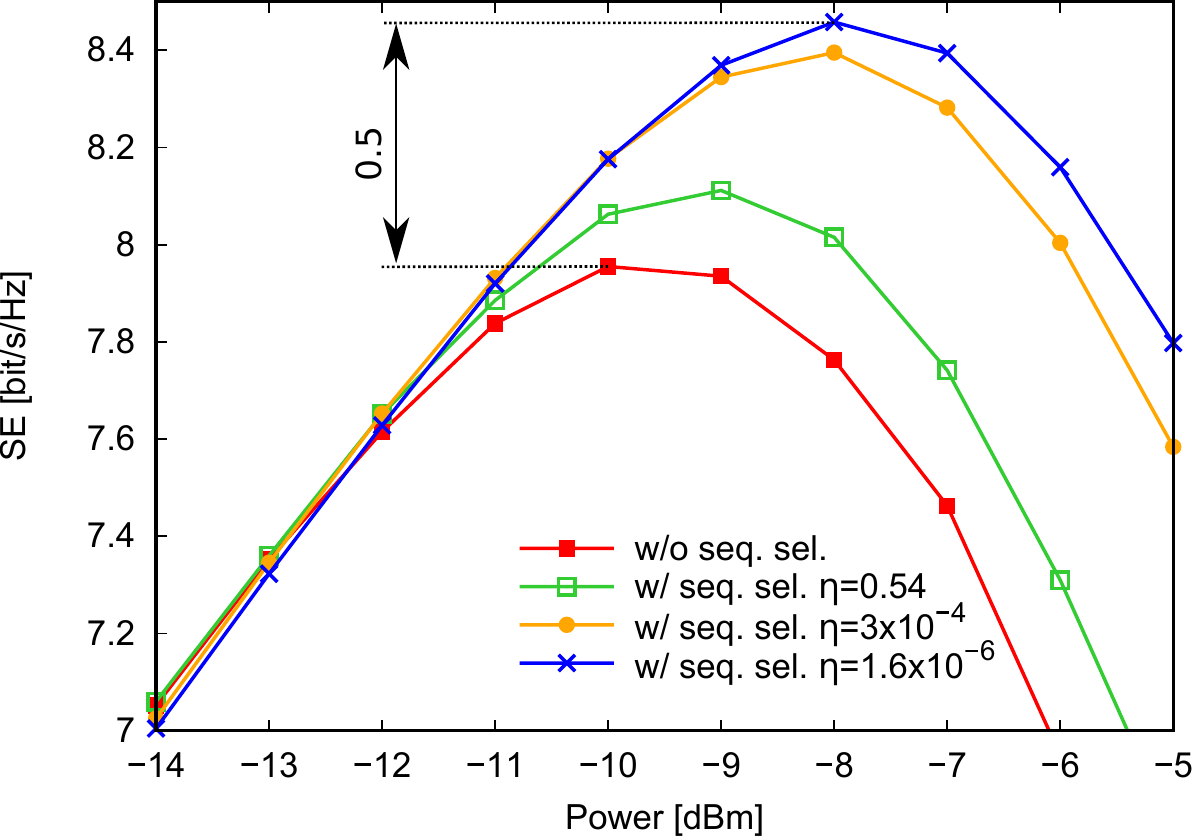}
\par\end{centering}
\hspace*{\fill}(a)\hspace*{\fill}\hspace*{\fill}(b)\hspace*{\fill}

\caption{\label{fig:SE_1pol-1ch}Achievable SE for the single-channel single-polarization
system with sequence selection: (a) SE versus acceptance rate for
different values of the block length $n$ and launch power; (b) SE
versus launch power for $n=256$ and different values of the acceptance
rate.}
\end{figure*}

As explained in Section~\ref{sec:sequence-selection}, by setting
a threshold on the cost function (the horizontal axis in Fig.~\ref{fig:pdf_cdf}),
it is possible to select only the ``good'' sequences that cause
a lower NLI; by reducing the threshold (moving it to the left), both
the acceptance rate (which equals the cdf) and the average NLI energy
decrease. The SE as a function of the acceptance rate is reported
in Fig.~\ref{fig:SE_1pol-1ch}(a) for different values of the block
length $n$ and two different launch powers: slightly more than the
optimal launch power without sequence selection (-9~dBm) and 1~dB
more. The SE increases monotonically for large $n$ (256 and 1024).
A similar behavior could be expected for $n=64$, given that the corresponding
cdf in Fig.~\ref{fig:pdf_cdf}(b) grows more slowly than $\sim\lambda^{n/3}$.
However, the theory in Section~\ref{sec:Analytical-results} and
the cost function (\ref{eq:cost_function_memoryless}) are based on
the assumption of a block-memoryless NLI, and do not account for inter-block
NLI. In fact, the value of the cost function computed for a specific
sequence by the above procedure depends also on the adjacent sequences,
meaning that a low-cost sequence selected in this way might, in fact,
have a higher cost when combined with different adjacent sequences.
The impact of inter-block NLI affects mostly the symbols at the edges
of the sequence, so that it becomes less and less relevant as $n$
increases. In principle, $n$ should be taken as large as possible
to obtain the best performance. However, as can be inferred from (\ref{eq:acceptance_rate_gamma})
and (\ref{eq:optimal_threshold}) and from Fig.~\ref{fig:pdf_cdf}(b),
a longer $n$ requires a lower acceptance rate to achieve the optimal
performance, which means testing many more sequences. The lowest acceptance
rate for which we were able to obtain reliable results ($N_{a}>1000$
accepted sequences) in this scenario is $\sim10^{-6}$. At this rate,
we have obtained the highest gain for $n=256$. However, a longer
$n$ should offer a higher potential gain, provided that a sufficiently
lower acceptance rate could be achieved. This is indeed suggested
by the behavior of the curve for $n=1024$. An alternative approach
to limit the impact of inter-block NLI without increasing $n$ is
the use of the more accurate cost function (\ref{eq:cost-function_AWGNmetric}),
as it will be shown in Section~\ref{subsec:WDM-dual-polarization-system}.

Fig.~\ref{fig:SE_1pol-1ch} also shows that without sequence selection
(or for a high acceptance rate) the lower launch power of -9~dBm
offers the best performance, as we are already beyond the optimal
power. However, when decreasing the acceptance rate, the SE curves
at higher power cross the ones at lower power, as the optimal launch
power increases. In general, to fully exploit the potential of the
sequence selection approach, it is necessary to simultaneously decrease
the acceptance rate and increase the launch power. In fact, by reducing
the acceptance rate, we select sequences that cause less NLI, so that
we can increase their launch power and hence improve the SNR.

Fig.~\ref{fig:SE_1pol-1ch}(b) shows the SE as a function of the
launch power for a block length $n=256$ and different values of the
acceptance rate. 
 As expected, reducing the acceptance rate improves the SE, not only
at the launch power at which the sequences have been selected, but
in the whole nonlinear regime. As a result, both the optimal launch
power and the peak SE increase by reducing the acceptance rate. Clearly,
a lower acceptance rate means also a higher rate loss, which explains
why the SE in the linear regime is slightly lower. The results in
Fig.~\ref{fig:SE_1pol-1ch}(a) and (b) suggest that an even higher
SE could be expected by further reducing the acceptance rate and increasing
the launch power for $n\ge256$.

\subsection{WDM Dual-Polarization System\label{subsec:WDM-dual-polarization-system}}

The second test is performed in the dual-polarization WDM scenario,
where each channel is affected by both inter- and intrachannel NLI.
In this case, we consider two different procedures to compute the
cost function and select the best sequences. The first one is the
fast procedure described in Section~\ref{subsec:Single-channel-single-polarizati},
simply extended to consider two polarizations. The second one, on
the other hand, considers the more accurate cost function (\ref{eq:cost-function_AWGNmetric})
to average out the impact of inter-block NLI (neglected by the first
approach) and will be hence referred to as \emph{averaged} procedure.
This should ensure that the cost of the selected sequences is actually
low, regardless of the adjacent sequences that are actually transmitted.
In both approaches, the cost function is evaluated in a single-channel
noiseless scenario, therefore preferring those sequences that cause
low intrachannel NLI, regardless of their impact in terms of interchannel
NLI and signal-noise interaction. The averaged procedure is implemented
by the following steps:
\begin{itemize}
\item generate a very long input sequence $\boldsymbol{x}^{N'}$ by drawing
samples from the unbiased distribution and evaluate the corresponding
output sequence $\hat{\boldsymbol{y}}^{N'}$ by SSFM propagation through
the noiseless channel;
\item take as proposed sequences the $N_{p}=N'/(n+N_{g})$ disjoint subsequences
$\{\boldsymbol{x}_{i(n+N_{g})+1}^{i(n+N_{g})+n}\}$, for $i=0,\ldots,N_{p}-1$,
that are obtained by dividing $\boldsymbol{x}^{N'}$ into subsequences
of length $n$, separated by $N_{g}$ guard symbols;
\item for each proposed sequence, compute the temporary cost function 
\begin{equation}
\lambda(\boldsymbol{x}_{i(n+N_{g})+1}^{i(n+N_{g})+n})=\left\Vert \hat{\boldsymbol{y}}_{i(n+N_{g})+1}^{i(n+N_{g})+n}-\boldsymbol{x}_{i(n+N_{g})+1}^{i(n+N_{g})+n}\right\Vert ^{2}
\end{equation}
\item repeat $N_{\mathrm{it}}$ times the previous steps by leaving the
proposed subsequences in $\boldsymbol{x}^{N'}$ unchanged, while randomly
changing all the guard symbols;
\item for each proposed sequence, compute the final cost function by averaging
the temporary cost functions over the $N_{\mathrm{it}}$ realizations;
\item accept the $N_{a}$ subsequences for which the final cost function
is below the desired threshold.
\end{itemize}
If more sequences are needed, the procedure is repeated several times.
Also in this case, the selection is performed at a single launch power.
The AIR and SE are then estimated on the central channel by following
the procedure described in Section~\ref{subsec:AIR-estimation},
where the sequences of length $N$ transmitted on the five WDM channels
are independently generated by concatenating in a random order the
selected subsequences of length $n$.

\begin{figure}
\begin{centering}
\includegraphics[width=1\columnwidth]{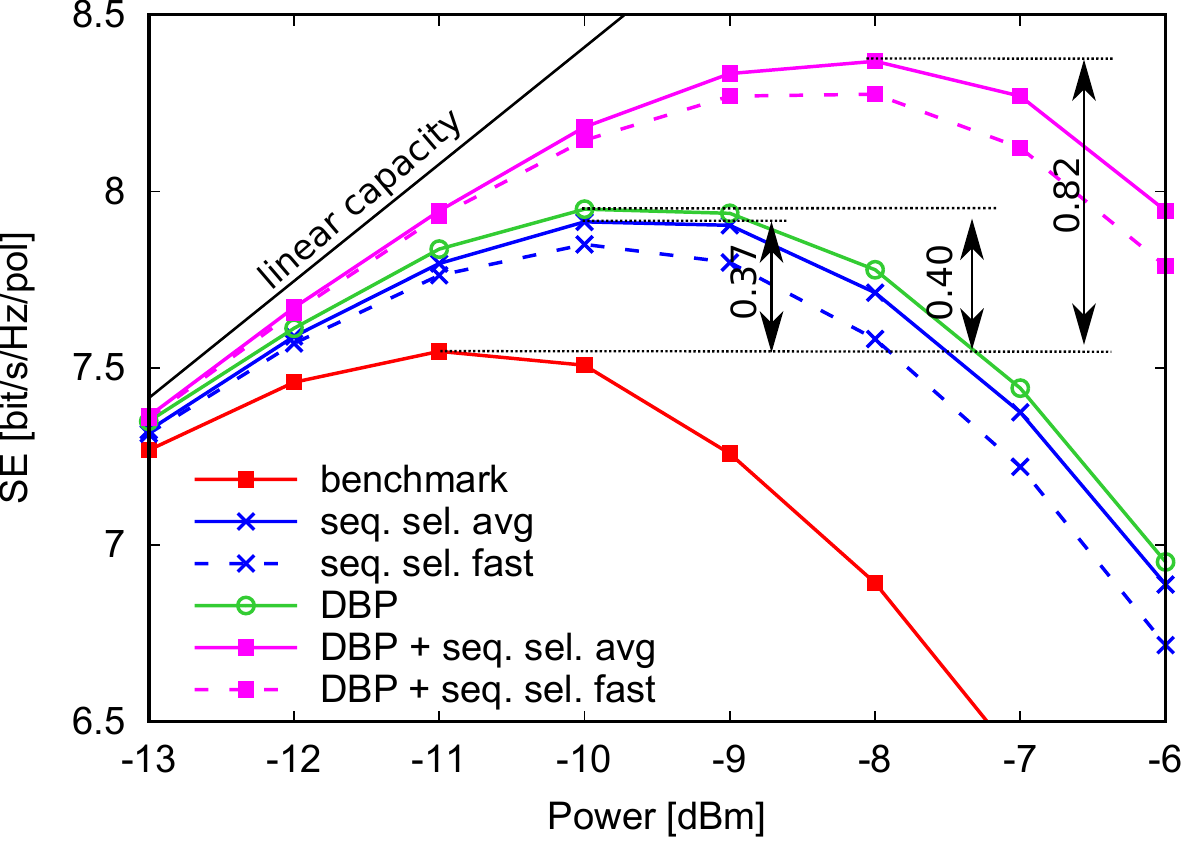}
\par\end{centering}
\caption{\label{fig:SE_5WDMchannels_2pol_AWGNdetection}Achievable SE for the
dual-polarization WDM system with sequence selection ($n=256$, $\eta=0.002$,
and two different cost functions), possibly combined with DBP.}
\end{figure}

Fig.~\ref{fig:SE_5WDMchannels_2pol_AWGNdetection} shows the SE obtained
in four different cases: the benchmark without sequence selection,
optimized for the linear regime; the system with sequence selection,
with $n=256$ dual-polarization symbols, $N_{p}=65512$ and $N_{p}=38350$
for the fast and averaged optimization, respectively, and $\eta=0.002$
(chosen as a good trade-off between performance and computational
complexity); the system with single-channel ideal DBP and without
sequence selection; and the system with both DBP and sequence selection,
where the same sequences selected without DBP are transmitted. 
 Sequence selection is performed by using either of the two described
procedures to estimate the cost functions (\ref{eq:cost_function_memoryless})
or (\ref{eq:cost-function_AWGNmetric}), the latter by considering
$N_{g}=800$ guard symbols and averaging over $N_{\mathrm{it}}=20$
realizations. As anticipated, the second cost function accounts for
the average impact of inter-block NLI, providing a more accurate selection
and a higher SE gain. With the improved cost function, sequence selection
alone yields an SE gain of 0.37~bits/s/Hz/pol with respect to the
benchmark. By comparison, DBP yields only a slightly higher gain of
0.39~bits/s/Hz/pol, while the combination of the two techniques yields
a higher total gain of 0.82~bits/s/Hz/pol. This means that the input
distribution provided by sequence selection, though optimized to reduce
intrachannel NLI, partly mitigates also interchannel NLI, so that
it provides an additional gain even when combined with DBP (which
completely removes intrachannel NLI).

\subsection{Combination of Shaping, DBP, and Improved Detection\label{subsec:Combination-of-shaping}}

As a final test, we combine the proposed sequence selection technique
for the optimization of the input distribution, with both DBP and
an improved decoding metric optimized for a channel with phase and
polarization noise (PPN). The aim is to maximize the achievable SE
to improve on the tightest available capacity lower bounds for the
WDM channel \cite{secondini2019JLT,garciagomez2021mismatched}. In
particular, by following \cite{secondini2019JLT}, we consider a random-walk
model for the phase and polarization evolution of the PPN metric,
and divide each WDM channel into four subcarriers to obtain the best
trade-off between the temporal and frequency coherence of the PPN
processes over each subcarrier. Moreover, by following \cite{garciagomez2021mismatched},
we optimize the power allocation per subcarrier to maximize the SE.
The SE with PPN decoding is estimated by using the general procedure
described in Section~\ref{subsec:AIR-estimation} and computing the
decoding metric by using the particle filtering approach described
in \cite{Dauwels:TrIT08,secondini2019JLT}.

\begin{figure}
\begin{centering}
\includegraphics[width=1\columnwidth]{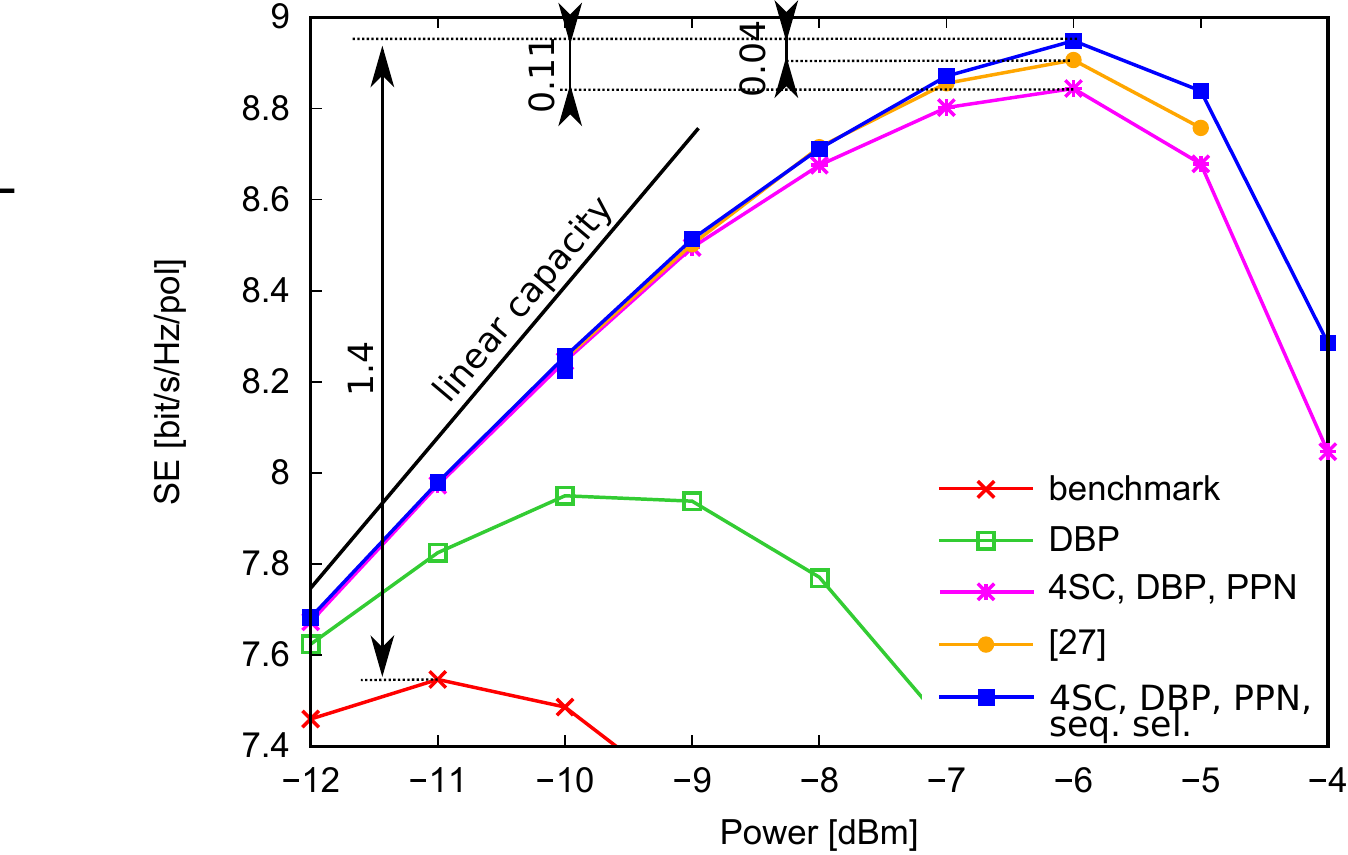}
\par\end{centering}
\caption{\label{fig:SE_5WDMchannels_2pol_4SC_PPNdetection}Achievable SE for
the dual-polarization WDM system with sequence selection ($n=512$,
$\eta=0.004$, averaged cost function) combined with single-channel
DBP, subcarrier multiplexing, per-subcarrier power optimization, and
PPN detection.}
\end{figure}

Fig.~\ref{fig:SE_5WDMchannels_2pol_4SC_PPNdetection} shows the new
SE lower bound obtained in this way (blue), compared to lower bound
\cite{garciagomez2021mismatched} (yellow) and to the SE obtained
with other configurations. Sequence selection is performed according
to the same procedure described in Section~\ref{subsec:WDM-dual-polarization-system}
and based on the cost function (\ref{eq:cost-function_AWGNmetric}),
considering $n=512$ dual-polarization symbols, $\eta=$0.004, $N_{p}=47820$,,
$N_{g}=800$, and $N_{\mathrm{it}}=20$. The only difference with
respect to Section~\ref{subsec:WDM-dual-polarization-system} is
that each proposed subsequence of $n$ symbols is formed by taking
$n/4$ symbols simultaneously transmitted on each subcarrier, and
different subsequences are separated by $N_{g}/4$ guard symbols on
each subcarrier. Therefore, as in Section~\ref{subsec:WDM-dual-polarization-system},
the resulting input distribution is optimized for the single-channel
noiseless scenario and does not account for the presence of interchannel
NLI, for the optimized power allocation per subcarrier, and for the
use of DBP and of an improved decoding metric. This allows to keep
the computational complexity of the selection process low, but provides
a suboptimal input distribution. Nonetheless, even in this case the
sequence selection approach provides an additional SE gain when combined
with DBP, PPN detection, and optimized power allocation, meaning that
the selected sequences are good also for this more complex scenario,
though not specifically optimized for it. In practice, DBP removes
intrachannel NLI, while both PPN detection and sequence selection
mitigate interchannel NLI.

The additional gain of 0.11~bit/s/Hz/pol provided by sequence selection
is significantly smaller than in the case studied in Section~\ref{subsec:WDM-dual-polarization-system},
but still sufficient to bring the overall gain to 1.4~bit/s/Hz/pol
with respect to the benchmark, surpassing by 0.04~bit/s/Hz/pol the
lower bound in \cite{garciagomez2021mismatched}. We remark that the
latter uses a slightly different model for PPN evolution; a whitening
filter to account also for the correlation of the additive noise;
and optimized delays between subcarriers and between WDM channels.
These differences explain the slightly higher SE of \cite{garciagomez2021mismatched}
(yellow curve) with respect to our SE without sequence selection (purple
curve). Nonetheless, the optimization of the input distribution based
on the sequence selection approach with the suboptimal cost function
(\ref{eq:cost-function_AWGNmetric}) is more effective than all these
techniques, providing a higher SE gain. We expect that our bound could
be only slightly improved by including also those techniques. On the
other hand, we expect that a more relevant SE gain should be obtained
by improving the selection procedure\textemdash e.g., by increasing
$n$ and $N_{\mathrm{it}}$, decreasing $\eta$, and tailoring the
cost function (\ref{eq:cost_function_conditional_entropy}) to the
WDM scenario and to the considered system configuration.

\section{Conclusion\label{sec:Conclusion}}

We have presented a new technique to optimize the input distribution
and lower-bound the capacity of the optical fiber channel in a very
general setting. The technique uses a rejection sampling algorithm,
driven by a properly defined cost function, to generate the sequences
of input symbols that are most suitable for the nonlinear channel
and the selected decoding metric. The corresponding AIR is then estimated
by Monte Carlo averaging. The rate loss induced by the selection process
is simply related to the acceptance rate of the rejection sampling
algorithm and removed from the AIR.

We have tested the proposed techniques in a few different scenarios.
In a simplified optical channel with block-memoryless NLI, the AIR
obtained with the selection process can be computed analytically,
showing that an unbounded growth of capacity with power is indeed
possible, provided that certain conditions on the distribution of
the NLI hold. In a still simple but more realistic single-polarization
single-channel scenario, numerical simulations demonstrate a significant
improvement of the AIR (and of the corresponding SE). The same simulations
also reveal that the above mentioned conditions on the NLI distribution
are practically met, suggesting that further AIR improvements are
possible by increasing the power and reducing the acceptance rate.
In a dual-polarization WDM scenario, the proposed technique yields
an AIR gain that is comparable to the gain offered by ideal single-channel
DBP; the gain is almost doubled when the two techniques are combined,
and further increases when also subcarrier multiplexing and a PPN
decoding metric are employed. The peak AIR obtained by the combined
techniques exceeds the highest capacity lower bound available in the
literature for the same channel, effectively establishing a new lower
bound. These results suggest that nonlinear constellation shaping,
if properly optimized in a high-dimensional space, can be a valid
replacement or an effective complement to DBP and other complex mitigation
strategies.

The definition of a suitable cost function is essential for the correct
optimization of the input distribution. We have proposed a general
definition, related to the minimization of the mismatched conditional
entropy; a more specific definition, suitable for intrachannel NLI
and an AWGN decoding metric; and a simplified definition, suitable
when intrachannel NLI is block-memoryless. Both the specific and the
simplified definitions have been implemented and tested numerically.
They work well in all the considered scenarios, even beyond the original
scope for which they are designed; in fact, the sequences selected
in this way mitigate also the impact of interchannel NLI, even when
combined with an improved decoding metric. The best performance is
obtained with the more complex cost function, whereas the simplified
one reduces significantly the computation cost of the selection procedure.

We expect that the numerical results and bounds obtained in this work
can be further improved by following different possible approaches.
The first one is merely a brute-force approach. The values used for
$n$, $N_{\mathrm{it}}$, and $\eta$ are not optimal, but simply
selected as a trade-off between performance and computational cost;
thus, better results can be obtained by employing more computational
resources to increase $n$ and $N_{\mathrm{it}}$ and to reduce $\eta$.
A more interesting approach is that of modifying the cost function.
In fact, for simplicity, the input sequences have been simply optimized
to minimize intrachannel NLI; higher gains are expected by better
tailoring the cost function to the relevant impairments and adopted
decoding metric, e.g., accounting also for interchannel NLI and for
the use of DBP and PPN decoding. An indication in this sense is provided
by (\ref{eq:cost_function_conditional_entropy}). The same approach
can be followed also to extend the application of the proposed techniques
to other channels and scenarios.

\section*{Acknowledgment}

The authors would like to thank Francisco Javier García-Gómez for
a useful discussion on the comparison between the capacity lower bounds
in this work and in \cite{garciagomez2021mismatched}.\bibliographystyle{IEEEtran}
\bibliography{ref}

\end{document}